\documentclass[12pt,preprint]{aastex}

\slugcomment{Revised 2004 September 29}

\shorttitle{Deep VLBI Imaging of Faint Radio Sources}
\shortauthors{Garrett, Wrobel \& Morganti}

\begin{document}

\title{Deep VLBI Imaging of Faint Radio Sources\\
       in the NOAO Bo\"otes Field}

\author{M.~A. Garrett\altaffilmark{1}, J.~M. Wrobel\altaffilmark{2},
and R.~Morganti\altaffilmark{3}}

\altaffiltext{1}{Joint Institute for VLBI in Europe, Postbus 2,
NL-7990 AA Dwingeloo, Netherlands; garrett@jive.nl}

\altaffiltext{2}{National Radio Astronomy Observatory, P.O. Box O,
Socorro, NM 87801; jwrobel@nrao.edu}

\altaffiltext{3}{Netherlands Foundation for Research in Astronomy,
Postbus 2, NL-7990 AA Dwingeloo, Netherlands; morganti@astron.nl}

\begin{abstract}
  We have conducted a deep, very long baseline interferometry (VLBI)
  observation at 1.4~GHz of an area of sky located within the NOAO
  Bo\"otes field, using the NRAO Very Long Baseline Array and 100-m
  Green Bank Telescope.  Applying wide-field VLBI techniques, a total
  of 61 sources, selected from a Westerbork Synthesis Radio Telescope
  (WSRT) image, were surveyed simultaneously with a range of different
  sensitivities and resolutions. The survey covered a total of
  1017~arcmin$^2$ $=$ 0.28~deg$^2$ divided into annular fields centered
  on $\alpha=$ 14h 29m 27.0s and $\delta=$ $+$35\arcdeg 28\arcmin
  30.00\arcsec\, (J2000). The inner 0-2\arcmin\, of the field reached
  an unprecedented 1 $\sigma$ rms noise level of $9~\mu$Jy~beam$^{-1}$
  and yielded two detections. VLBI J142923.6466 and VLBI J142934.7033
  have brightness temperatures in excess $10^{6}$~K and locate the
  active nucleus of their host galaxies (NDWFS J142923.6$+$352851 and
  NDWFS J142934.7$+$352859 with $I \sim 16.3^{m}$ and $I \sim 19.6^{m}$
  respectively). Further deep surveys of the inner 2-4\arcmin\, and
  4-6\arcmin\, of the field, with 1 $\sigma$ rms noise levels of
  11-19~$\mu$Jy~beam$^{-1}$, detected a previously known source, VLBI
  J142910.2224, a quasar with a brightness temperature in excess of
  $10^{9}$~K that was also used during these observations as an in-beam
  phase calibrator. The deep VLBI survey between 0\arcmin\, and
  6\arcmin\, thus detected 3 radio sources, drawn from a total of 24
  targets. A shallower VLBI survey, conducted between 6\arcmin\, and
  18\arcmin\, of the field center, and with 1 $\sigma$ rms noise levels
  of 37-55~$\mu$Jy~beam$^{-1}$, detected a further 6 radio sources,
  drawn from 37 additional targets. Each of those 6 VLBI detections has
  a brightness temperature in excess of $10^5$~K; this hints that those
  6 are accretion-powered, a suggestion reinforced by the double
  structure of 3 of them. Combining both the deep and shallow VLBI
  surveys, optical identifications are available for 8 of the 9 VLBI
  detections.  Only VLBI J142906.6095 remains unidentified ($I >
  25.6^{m}$), quite unusual as its integrated WSRT flux density is
  20~mJy. Two other sources are not detected in K-band ($K > 18.5^{m}$)
  suggesting that some significant fraction of these compact radio
  sources may be located at $z > 1$. The VLBI detection rate for
  sub-mJy WSRT radio sources is 8$^{+4}_{-5}$\%.  The VLBI detection
  rate for mJy WSRT sources is higher, 29$^{+11}_{-12}$\%. This
  observational trend is expected from a rapidly evolving radio source
  population.  Moreover, this trend supports deep radio surveys, at
  lower resolution, that infer the radio emission associated with
  fainter sub-mJy and microJy sources arises via processes associated
  with extended regions of star formation. The 9 VLBI detections
  reported here pin-point the precise location of active nuclei or
  candidate active nuclei, and their VLBI positions can help to anchor
  the NOAO Bo\"otes field to the International Celestial Reference
  Frame.  The simultaneous detection of several sub-mJy and mJy radio
  sources, in a single observation, suggest that their combined
  response may be used to self-calibrate wide-field VLBI data.  There
  is every prospect that future deep VLBI observations can take
  advantage of this wide-field technique, in any random direction on
  the sky, thereby generating large-area, unbiased surveys of the faint
  radio source population.
\end{abstract}


\section{Motivation}

Evolutionary models of extragalactic radio sources
\citep[e.g.,][]{con84} indicate that the overwhelming majority of
sources with integrated flux densities in excess of $S_{\rm I} \sim
1$~mJy are typically identified with active galaxies, energised by
accretion onto massive black holes.  These bright sources have a
median angular size of 5-10\arcsec\, \citep[e.g.,][]{bon03,con03} and
are usually resolved by the VLA on arcsecond and sub-arcsecond scales.
Often a significant fraction of the total radio emission resides
within a compact and unresolved radio ``core'', with a size measured
on milliarcsecond (mas) and sub-mas scales using the techniques of
very long baseline interferometry (VLBI).

Fainter sub-mJy and microJy radio sources have a higher probability of
being energised by, and coextensive with, star formation in galaxies
located at moderate redshifts $z \sim 0.3-1$ \citep{ric00,sma02}.
\citet{gar02} has shown that the bulk of these distant systems obey the
locally derived \citep[e.g.,][]{con92} FIR-radio correlation.  Indeed,
only about 20\% of the sub-mJy radio source population appears to be
optically identified with active galaxies; a further 10\% are optically
faint sources with $I > 25^{m}$ and are difficult to classify from
their weak radio emission alone \citep{mux04}. MERLIN-VLA observations
of faint radio sources in the HDF show that the microJy radio sources
are resolved and typically sub-galactic in their overall extent, with
typical measured sizes of 0.2-3 arcseconds \citep{mux04}. This is
consistent with the analysis of \citet{con92}, limiting the maximum
brightness temperature of star forming galaxies to $< 10^{5}$~K at
frequencies above 1 GHz.

A recent study conducted by \citet{hai04} suggests that a significant
fraction of the faint radio source population may be located at high
redshift. At the level of 10~$\mu$Jy they predict, in their simplest
models, source surface densities of 10~deg$^{-2}$ at $z > 10$. However,
at these faint sensitivity levels it is increasingly difficult to
identify radio-loud active galaxies from the dominant star forming
radio source population, especially in the case of optically faint
systems.  High-resolution VLBI observations resolve out these extended
star forming galaxies and, currently, any compact radio sources in
these galaxies, such as luminous Type IIn SNe
\citep[e.g.,][]{wil02,smi98}, would be too faint to detect at
cosmological distances.  On the other hand, VLBI is very well matched
to detect the very compact radio emission associated with the
relativistic outflows generated by accretion onto massive black holes
in active galaxies.

High sensitivity VLBI observations, covering a large fraction of sky,
thus provide a simple and direct method of identifying these faint, and
possibly distant, radio-loud active galaxies.  Free from the effects of
dust obscuration, deep and wide-field VLBI studies can therefore contribute
to the cosmic census of active galaxies (and their energising massive
black holes), and together with redshift information, can potentially
probe the accretion history of the early Universe.  In addition, the
positions of the compact VLBI detections can be measured very
accurately, via the use of standard phase-referencing techniques.
Astrometric precision at the mas level can be achieved routinely, and
cross identification with sources detected at other wave-bands can
anchor these non-radio observations to the International Celestial
Reference Frame (ICRF). In crowded and deep fields, such precision
astrometry may be useful in identifying the counterparts of faint or
obscured sources at other wavelengths. 

The observations presented here build on earlier wide-field VLBI
studies of the Hubble Deep Field North (HDF-N), conducted by
\citet{gar01} using the European VLBI Network (EVN). Unlike traditional
snapshot VLBI surveys, the wide-field approach permits many potential
targets in the field to be imaged simultaneously, taking full advantage
of the sensitivity and the coverage in the {\em (u,v)} plane associated
with the total duration of the observations and the characteristics of
the array.  Using this observing strategy, it is possible to survey
many sources (as in the case of snapshot observations) but with much
greater sensitivity.

In this paper, we present the results of a deep VLBI survey that covers
a region of 1017~arcmin$^2$ $=$ 0.28~deg$^2$, surveying 61 potential
sub-mJy and mJy radio source targets at angular resolution ranging from
about 10 to 100~mas. For sources located at cosmological distances, the
linear resolution corresponding to 10~mas is 90~pc or finer for a {\em
  WMAP\/} cosmology \citep{spe03}. The VLBI observations and
correlation are described in section~\ref{obs}, while section~\ref{cal}
describes the VLBI calibration and imaging.  Definition of the VLBI
survey fields, survey depths, and target-source selection appear in
section~\ref{sel}.  Results of the VLBI survey are presented in
section~\ref{res} and discussed in section~\ref{dis}.  The paper closes
in section~\ref{fut} with a synopsis of the future prospects for deep,
wide-field VLBI surveys.

\section{Observations and Correlation}
\label{obs}

\subsection{VLBA+GBT Observations}
\label{vlbisec}

Observations at 1.4~GHz of a region of the NOAO-N Bo\"otes field were
made with the NRAO Very Long Baseline Array (VLBA) and Green Bank
Telescope (GBT), recording at a data rate of 256~Mbit~s$^{-1}$ across
a 64-MHz band.  For descriptions of the VLBA and GBT, see
\citet{nap94} and \citet{jew00}, respectively.  The observations
spanned three 8-hour segments on 2002 August 3, 4, and 5 UT.  The
pointing position for the VLBI array was chosen as $\alpha=$ 14h 29m
22.0s and $\delta=$ $+$35\arcdeg 28\arcmin 50.00\arcsec\, (J2000) and
included an area of sky surveyed as part of an earlier VLBA snapshot
survey of the region \citep{wro04}.  In particular, with this pointing
position, the compact 20-mJy source J142910.223$+$352946.86 fell
within the FWHM of the primary beam of all the antennas, including the
100-m GBT.  Figure~1 presents the VLBA+GBT antenna pointing position
and the in-beam calibrator, superimposed upon the Westerbork Synthesis
Radio Telescope (WSRT) image of the field described in section
\ref{wsrtsec}.

A combination of nodding and in-beam phase referencing was employed
\citep{wro00},\footnote{Available at
  www.aoc.nrao.edu/vlba/html/MEMOS/scimemos.html.}, resulting in an
integration time on the target field of 20 hours.  J1426$+$3625 was
used as the conventional phase calibrator and the VLBI astrometry
reported in this paper assumes a position for this calibrator of
$\alpha=$ 14h 26m 37.0874939s and $\delta=$ $+$36\arcdeg 25\arcmin
9.573930\arcsec\, (J2000) in the ICRF (Extension 1) \citep{ier99,ma98}.
Both the observations and correlation processing assumed a coordinate
equinox of 2000. Two correlation passes were made within the GBT
primary beam of FWHM 8.6\arcmin.  One correlation pass was centered on
the VLBA position of the in-beam calibrator J142910.223$+$352946.86
given in Table~3.  Another correlation pass was centered on $\alpha=$
14h 29m 27.0s and $\delta=$ $+$35\arcdeg 28\arcmin 30.00\arcsec\,
(J2000), chosen because of the favorable number of potential targets
that lay within a radial extent of 2\arcmin.  The various VLBI survey
fields introduced in section \ref{sel} are cast in terms of their
radial extents with respect to this correlation phase center.

In order to reduce the effects of bandwidth and time smearing, and
thus to image out as large a field of view as possible, the VLBA
correlator generated data with 1024 spectral points per baseline and
an integration time of 0.524~s.  The VLBA station at Mauna Kea, HI,
was omitted from the correlation process, due to the physical
limitation on the total output data rate.  With a correlator output
data rate of 1~Mbyte~s$^{-1}$, a final data set size of 60~Gbytes was
realised for each of the two correlator passes.

\subsection{WSRT Observations}
\label{wsrtsec}

Previous WSRT observations of the Bo\"otes field \citep{dev02} also
encompassed the region of sky observed by the VLBA+GBT.  In addition,
a further 12-hour, 1.4-GHz observation of this particular region of
the Bo\"otes field was conducted \citep{mor02}, using the full 160 ($8
\times 20$) MHz WSRT observing band and employing the default
continuum frequency set-up (with the eight bands centered between
1.311 and 1.450 GHz).  A short 20-minute scan on 3C~48 was used to
initially calibrate the data.  For each of the 8 bands, 64 spectral
channels were generated (a total of 512 spectral points were obtained
for the 160 MHz band) and 4 polarization products were recorded.  The
data were calibrated and images generated using the MIRIAD software
package \citep{sau95}.  The \citet{mor02} WSRT image of the field,
after primary beam correction, is shown in Figure~1.  The image
reaches a 1 $\sigma$ rms noise level of $\sim 14 ~\mu$Jy~beam$^{-1}$
in this area.  The AIPS task {\sc JMFIT} was used to determine the
positions of regions of radio emission in the WSRT image, as well as
their peak flux densities $S_{\rm P}(WSRT)$ and integrated flux
densities $S_{\rm I}(WSRT)$.  WSRT sources that met the selection
criteria defined in section \ref{sel} and Table~1 were added to the
lists of potential target sources for the VLBA+GBT observations; those
lists appear in Tables~2-7.  The WSRT astrometry is typically accurate
to $\sim 1-2\arcsec$.  This estimate is consistent with the measured
positions of the WSRT sources detected during the VLBA+GBT
observations (see Tables~2-7).  In Figure~1, the positions of
potential radio source targets for the VLBA+GBT observations are
indicated by a cross.

\section{VLBI Calibration and Imaging}
\label{cal}

The data from the in-beam calibrator pass were edited, averaged, and
calibrated in the NRAO AIPS package, and the gain parameters applied to
the unaveraged, target-pass data.  The response of the in-beam
calibrator was also subtracted from the target {\em (u,v)} data.  The
AIPS task {\sc IMAGR} was used to make naturally weighted dirty images
and beams of the target regions selected from the WSRT image.  Targets
falling within a certain radial distance were imaged simultaneously
using the multi-field option within {\sc IMAGR}.  Given the astrometric
accuracy of the WSRT, together with the fact that many of the target
sources may have extended radio structures on arcsecond scales, a large
possible error between the WSRT and VLBI positions was assumed: each
dirty image therefore subtended a square of at least 6\arcsec\, on a
side.  For each target, three such dirty images were made,
corresponding to the three epochs of observation.  The three images for
each target source were then simply co-added (averaged) to form the
final target image. We estimate that the survey analysis required two
orders of magnitude more processing power and disk storage requirements
than a typical (single target) VLBI observation. The amount of
processing power required does not only depend on the data set size
(e.g. 120 Gbytes of data for this observation c.f. $\sim 5$Gbyte for a
typical VLBI continnum observation) but also the number of target
images that must be generated ($\sim 60$ target images in this
experiment c.f. 1 target image for typical VLBI experiments).

The largest VLBI images of the targets contained about 400,000
synthesized-beam areas.  Hence, the chances of misidentifying a random
noise spike as a real detection are significant.  In order to avoid
spurious detections, the dirty images of the target regions were
inspected for valid sources using a conservative 6 $\sigma$ detection
threshold.  Most of the sources we detected lie well above this
threshold and were therefore bright enough to benefit from the
application of the CLEAN deconvolution algorithm.  Cleaned images were
produced using the AIPS task {\sc APCLN}.  Visibility-based cleaning is
currently prohibitively expensive in terms of computing requirements.

\section{Survey Fields, Survey Depths, and Source Selection}
\label{sel}

We split the survey into 6 separate fields, demarcated by increasing
radial distance from the correlation phase center.  These are referred
to as the 0-2\arcmin, 2-4\arcmin, 4-6\arcmin, 6-9\arcmin, 9-12\arcmin,
and 12-18\arcmin\, survey fields. We also refer to deep and shallow
survey fields.  The deep survey fields (0-2\arcmin, 2-4\arcmin\, and
4-6\arcmin) attains 1 $\sigma$ rms noise levels of
9-19~$\mu$Jy~beam$^{-1}$, significantly better than the $\sim
45~\mu$Jy~beam$^{-1}$ for a typical 8-hour VLBA observation.  The
shallow survey field (6-9\arcmin, 9-12\arcmin\, and 12-18\arcmin)
attains 1 $\sigma$ rms noise levels of 37-55~$\mu$Jy~beam$^{-1}$,
comparable with a typical 8-hour VLBA observation.  The measured noise
levels are in good agreement with those expected from thermal noise
considerations.

Our survey attempts to detect sources at large radial distances from
the antenna pointing position and the correlation phase center.  The
fall-off of the response of the primary beam is one effect that limits
the depth of each survey field.  In addition, the effect of employing a
restricted {\em (u,v)} range in most survey fields, in order to reduce
the effect of bandwidth and time smearing, leads to a noise level that
also increases with radial distance from the correlation phase center.

In particular, the reduced response in the VLBA+GBT images presented
here is composed of 4 independent components: the GBT primary beam
response ($R_{GBT}$), the VLBA primary beam response ($R_{VLBA}$) and
the reduced response due to time and bandwidth smearing effects
($R_{t}$, $R_{bw}$).  The combined reduced response, $R$, is given by:

\begin{equation}
R = R_{t} R_{bw} \sqrt{R_{GBT}R_{VLBA}}
\end{equation}

The GBT only contributes to the 0-2\arcmin\, and 2-4\arcmin\, surveys,
due to the sharp fall off of the primary beam response beyond its
half-power point.  For the 4-6\arcmin, 6-9\arcmin, 9-12\arcmin, and
12-18\arcmin\, survey fields, the reduced response is given by:

\begin{equation}
R = R_{t} R_{bw} R_{VLBA}
\end{equation}

We have estimated $R_{t}$ and $R_{bw}$ following \citet{Bri99}.  For
the VLBA 25-m antennas, a fiducial estimate of the primary beam at FWHM
is 29\arcmin, but a functional form for the measured primary beam
response is not available.  Since the primary beam response of the VLBA
25-m antennas is expected to be similar to that for the VLA 25-m
antennas, we have adopted for the VLBA antennas the fitted function
documented in the AIPS task {\sc PBCOR} to model the response of the
VLA primary beam.  For the GBT, we have assumed a pure Gaussian
response (F.~Ghigo, 2004, private communication) with a FWHM of
8.6\arcmin.  In Table~1, we calculate the total response at the outer
edge of each of the main survey fields.  For the naturally weighted
images associated with the inner 0-2\arcmin\, and 2-4\arcmin\, survey
regions, we assume that the longest GBT baseline (3326 km to the VLBA
station at Brewster, WA), is the appropriate limiting factor for the
time and bandwidth smearing response.

For any given region we selected target sources with measured peak
flux densities in the WSRT image, $S_{\rm P}(WSRT)$, that satisfied
the following constraint:

\begin{equation}
S_{\rm P}(WSRT) >  \frac{6\sigma_{noise}}{R}
\end{equation} 

Details of the properties of the various survey fields, including the
survey depth, are presented in Table~1.  The quoted survey depth is
conservative, in the sense that the value of $R$ is determined from the
maximum radial extent encompassed by each survey field; and for the
4-6\arcmin, 6-9\arcmin, 9-12\arcmin, and 12-18\arcmin\, fields, from
the maximum contributing baseline of the VLBI array.  The deep survey
is capable of detecting sub-mJy radio sources over its full extent of
254~arcmin$^2$ $=$ 0.07~deg$^2$.  The shallower and lower-resolution
survey covers an additional 763~arcmin$^2$ $=$ 0.21~deg$^2$ of sky,
targeting the brighter sub-mJy and mJy radio sources in the field.  The
resolution at FWHM of the deep and shallow surveys range between 7 and
26 mas, and 27 and 115 mas, respectively.

\section{Results}
\label{res}

Details of the target sources and VLBI detections are presented in
Tables 2-7 for the various deep and shallow fields. The WSRT flux
densities are corrected for primary beam attenuation and the absolute
flux density scale is estimated to be better than 2\%. The absolute
flux density scale of the VLBI observations are expected to be better
than 5\%. The VLBI flux densities have not been corrected for primary
beam attenuation, the correction factors scale as $1/R$ as detailed in
Table 1 for the various survey regions. We estimate the error in VLBI
positions in the 0-2\arcmin, 2-4\arcmin \& 4-6\arcmin\, fields to be
set by errors introduced by the ionosphere. With a separation of
1.2\arcdeg between our primary phase reference and the target fields,
we estimate an error of $\sim 1-2 $ mas in each coordinate. In fact we
measure a $\sim 2$~mas offset in each coordinate between the 1.4 GHz
position for the in-beam phase reference VLBI J142910.2224 (target 11)
presented in this paper, and the 5~GHz position for the same source
measured by \cite{wro04}.  This offset is consistent with our estimated
absolute VLBI position error, given that the source is likely to
present slightly different radio source structures at 1.4 and 5~GHz. In
the (heavily tapered) lower-resolution 6-9\arcmin, 9-12\arcmin \&
12-18\arcmin\, fields, r.m.s. noise errors limit the positional
accuracy. Since all sources are detected above a $6-\sigma$ detection
threshold, the error in each coordinate for these outer fields is
expected to be better than $\sim 4$~mas for the 6-9 arcmin field, and
$\sim 10$~mas in the 9-12\arcmin \& 12-18\arcmin, outer fields.

\subsection{The Deep 0-2\arcmin\, Survey}

The 0-2\arcmin\, survey is the deepest one listed in Table~1, since
there are no restrictions on the {\em (u,v)} data used to produce VLBI
images of the targets.  For this survey field we identified 10
potential target regions with peak flux densities $S_{\rm P}(WSRT) >
~74~\mu$Jy (Table~2).  Eight of the target regions showed no positive
(or negative) peaks above (or below) the 6 $\sigma$ detection
threshold.  Two of the target regions contained positive peaks at the
34 $\sigma$ and 31 $\sigma$ level.  These are Target 3 and Target 10,
discussed below.  Naturally-weighted images of these VLBI detections
are shown in Figures~2 and 3. Figures~4 and 5 present uniformly
weighted images of the same sources (robustness -4).  In order to
maximise our chances of detecting extended VLBI sources, images of all
10 targets were made at lower resolution, using reduced {\em (u,v)}
ranges from zero to 12M$\lambda$, 6M$\lambda$, and 3M$\lambda$.  No
additional targets were detected, nor was extended emission detected
from Target 3 or Target 10.

\subsubsection{Target 3. $S_{\rm I}(WSRT) \sim 6.1$~mJy}

VLBI J142923.6466 locates the active nucleus within the optical host,
NDWFS J142923.6$+$352851, an elliptical galaxy with total $I \sim
16.3^{m}$ and $K \sim 13.9^{m}$ \citep{jan04}.  Figure~2 was analysed
with AIPS task {\sc JMFIT} to yield the position, peak flux density
$S_{\rm P}(VLBI)$, and integrated flux density $S_{\rm I}(VLBI)$
entered in Table~2.  Since $S_{\rm I}(VLBI) \sim 0.46$~mJy, less than
ten percent of the integrated WSRT flux density has been recovered.  A
lower limit for the brightness temperature of the VLBI component of the
source is $4.5 \times 10^{6}$~K, suggesting this WSRT source is probably
a radio galaxy. Assuming the strong correlation between K-band
magnitude and the redshift of luminous radio galaxies also applies to
fainter, less luminous systems (the $K-z$ relation e.g. \citet{jar01}),
we estimate $z \sim 0.2$ for this source. At this redshift 10 mas
corresponds to 33 pc.

\subsubsection{Target 10. $S_{\rm I}(WSRT) \sim 0.66$~mJy}

Figure~5 shows that VLBI J142934.7033 is resolved into a double
separated by about 5~mas.  The integrated flux density is $S_{\rm
  I}(VLBI) \sim 0.46$~mJy (Fig.~3, Table~2), so most of the integrated
WSRT flux density has been recovered.  A lower limit for the brightness
temperature of the VLBI component of the source is $2.8 \times
10^{6}$~K.  Moreover, this north-south double is misaligned with its
east-west host galaxy, NDWFS J142934.7$+$352859, which has a total $I
\sim 19.6^{m}$ and $K \sim 16.3^{m}$ \citep{jan04}.  This radio-optical
misalignment strongly suggest that this sub-mJy WSRT source is
energised by a black hole rather than by star formation.  Assuming the
$K-z$ relation applies, we estimate $z \sim 0.6 $, for which 10 mas
corresponds to 67 pc.

\subsubsection{Target 9. $S_{\rm I}(WSRT) \sim 1.7$~mJy}

One of the brightest potential targets in the 0-2\arcmin\, survey
region, this WSRT source is identified with the disk galaxy NGC\,5646
at $z \sim 0.03$ \citep{dev02}.  This target is not detected with VLBI
at resolutions of 10-100 mas (~6-60~pc).  Figure~6 presents the full
resolution image of the central NGC\,5646 target region.  Indeed, for
this source, several regions each spanning 6\arcsec\, ($\sim 3.6$ kpc)
on a side, were placed across the extended WSRT source, in case
luminous SNe or SNR might be detectable.  This VLBI non-detection,
together with the fact that the WSRT emission follows the optical
isophotes \citep{mor02} and obeys the FIR-radio correlation
\cite{con02}, strongly supports a star formation origin for this mJy
WSRT source.

\subsection{The Deep 2-4\arcmin\, Survey}

Within the 2-4\arcmin\, survey field, we formally identified 3
potential targets with WSRT peak flux densities in excess of $\sim
0.14$~mJy (Table~3).  Two of the target regions showed no positive (or
negative) peaks above (or below) the 6 $\sigma$ detection threshold.
The remaining target, Target 11, is discussed below.  Lower resolution
images were made of all 3 targets, using reduced {\em (u,v)} ranges
from zero to 6M$\lambda$ and 3M$\lambda$.  No additional source
detections were made and no extended emission (from Target 11) was
detected.

\subsubsection{Target 11. $S_{\rm I}(WSRT) \sim 25$~mJy}

VLBI J142910.2224 has previously been detected by \citet{wro04}, who
validated its suitability as an in-beam calibrator for this study.  The
naturally weighted image of the source, obtained via the initial data
analysis of one of the three VLBI observations, recovers essentially
all of the integrated WSRT flux density (Fig.~7, Table~3).  Identified
as a quasar of unknown redshift, VLBI J142910.2224 is identified with
NDWFS J142910.2$+$352946 ($I \sim 18.4^{m}$) \citep{jan04}. A lower
limit for the brightness temperature of the VLBI component of the
source is $1.5 \times 10^{9}$~K. This mJy WSRT source must be dominantly
energised by a super-massive black hole.

\subsection{The Deep 4-6\arcmin\, Survey}

Within the 6-12\arcmin\, survey field, we identified 11 potential
targets with peak flux densities in excess of $\sim 0.23$~mJy
(Table~4).  None of the 11 target regions showed positive (or negative)
peaks above (or below) the 6 $\sigma$ detection threshold.  Lower
resolution images were also made of the target regions, using reduced
{\em (u,v)} ranges from zero to 6M$\lambda$ and 3M$\lambda$, but no
targets were detected.

\subsection{The Shallow 6-9\arcmin\, Survey}

Within the 6-9\arcmin\, survey field, we identified 9 potential
targets with peak flux densities in excess of $\sim 0.36$~mJy
(Table~5).  Eight of the target regions showed no positive (or
negative) peaks above (or below) the 6 $\sigma$ detection threshold.
Only one target, Target 33, contained a positive peak at the 7
$\sigma$ level. 

\subsubsection{Target 33. $S_{\rm I}(WSRT) \sim 0.44$~mJy}

VLBI J143002.5631 is not resolved and has an integrated flux density
$S_{\rm I}(VLBI) \sim 0.51$~mJy (Fig.~8, Table~2), recovering all of
the integrated WSRT flux density.  The host galaxy NDWFS
J143002.5$+$353035 has a total $I \sim 16.5^{m}$ and $K \sim 17.0^{m}$,
and may have a blue companion to the south-west \citep{jan04}.  A lower
limit for the brightness temperature of the VLBI component of the
source is $1.7 \times 10^{5}$~K.  If this sub-mJy WSRT source is a
radio galaxy, then the $K-z$ relation suggests $z \sim 0.8$, for which
10 mas corresponds to 75 pc.

\subsection{The Shallow 9-12\arcmin\, Survey}

Within the 9-12\arcmin\, survey field, we identified 17 potential
targets with peak flux densities in excess of $\sim 0.5$~mJy
(Table~6).  Fifteen of the target regions showed no positive (or
negative) peaks above (or below) the 6 $\sigma$ detection threshold.
Two of the target regions contained positive peaks at levels of 13
$\sigma$ for Target 34 and 8 $\sigma$ for Target 35.  Both are
discussed below.

\subsubsection{Target 34. $S_{\rm I}(WSRT) \sim 7.5$~mJy}

VLBI J142835.5389 is resolved into a double with a combined flux
density of $\sim 2.6$~mJy (Fig.~9, Table~6).  More than 30\% of the the
integrated WSRT flux density has been recovered.  The host galaxy NDWFS
J142835.5$+$353154 has a total $I \sim 22.8^{m}$ \citep{jan04}. The
source is not detected in K-band ($K > 18.5^{m}$ for a 50\% completeness
limit) suggesting it may be located at $z > 1$.  A lower limit for the
brightness temperature of the VLBI component of the source is $1.5
\times 10^{6}$~K suggesting that this mJy WSRT source is probably a
radio galaxy.

\subsubsection{Target 35. $S_{\rm I}(WSRT) \sim 1.0$~mJy}

VLBI J142835.9570 is not resolved and has an integrated flux density
$S_{\rm I}(VLBI) \sim 0.46$~mJy (Fig.~10, Table~6) that recovers at
least 40\% of the integrated WSRT flux density. 
A lower limit for the
brightness temperature of the VLBI component of the source is $3
\times 10^{5}$~K. 
The host galaxy NDWFS J142835.9$+$352537 has a total $I \sim
21.2^{m}$ and $K \sim 18.3^{m}$ \citep{jan04}.  If this sub-mJy
WSRT source is a radio galaxy, then the $K-z$ relation suggests $z > 1$.

\subsection{The Shallow 12-18\arcmin\, Survey}

Within the 12-18\arcmin\, survey field, we identified 17 potential
targets with peak flux densities in excess of $\sim 1.2$~mJy
(Table~7).  Fourteen of the target regions showed no positive (or
negative) peaks above (or below) the 6 $\sigma$ detection threshold.
Three of the target regions contained positive peaks at levels of 18
$\sigma$ (Target 52), 18 $\sigma$ (Target 55), and 6 $\sigma$ (Target
58), and are discussed below.

\subsubsection{Target 52. $S_{\rm I}(WSRT) \sim 10$~mJy}

Figure~11 shows that VLBI J142842.5476 is resolved into a double.  With
$S_{\rm I}(VLBI) \sim 5.4$~mJy (Fig.~11, Table~7), at least half of the
integrated WSRT flux density has been recovered. A lower limit for the
brightness temperature of the VLBI component of the source is $5 \times
10^{5}$~K. Moreover, this double is misaligned with its elongated host
galaxy, NDWFS J142842.5$+$354327, which has a total $I \sim 19.9^{m}$
and $K \sim 16.4^{m}$ \citep{jan04}.  This radio-optical misalignment
strongly suggests that this mJy WSRT source is energised by a black
hole rather than by star formation.  The $K-z$ relation would then
suggest $z \sim 0.6$.

\subsubsection{Target 55. $S_{\rm I}(WSRT) \sim 20$~mJy}

Figure~12 presents VLBI J142906.6095. With $S_{\rm I}(VLBI) \sim
13$~mJy (Fig.~12, Table~7), much of the integrated WSRT flux density is
recovered. A lower limit for the brightness temperature of the
VLBI component of the source is $2.7 \times 10^{6}$~K. The optical
counterpart must be fainter than $I \sim 25.6^{m}$ \citep{bro03,jan04},
quite unusual for such a strong radio source. There is no detection at
K-band ($K > 18.5^{m}$) suggesting the source may be located at $z > 1$.

\subsubsection{Target 58. $S_{\rm I}(WSRT) \sim 2.0$~mJy}

Figure~13 shows that VLBI J142941.6843 is resolved into a double.  With
$S_{\rm I}(VLBI) \sim 0.87$~mJy (Fig.~13, Table~7). A lower limit for the
brightness temperature of the VLBI component of the source is $1.3
\times 10^{5}$~K.  A candidate host galaxy, NDWFS J142941.8$+$351257,
has a total $I \sim 23.3^{m}$ \citep{jan04}. There is no detection at
K-band ($K > 18.5^{m}$) suggesting the source may be located at $z > 1$.

\section{Discussion}
\label{dis}

This paper presents the first attempt to characterise the sub-mJy and
mJy radio source population at mas resolutions.  A total of 61
potential targets were selected based on their peak WSRT flux
densities, and 9 were detected at or above the 6 $\sigma$ rms noise
level in the VLBI images. Most of the sources are probably located at
moderate redshifts ($z \sim 0.1-1$) but the non-detection of two
objects in K-band and one object in both I and K-band, suggests some
significant fraction of these faint, compact sources may be located at
$z >1$. Our observations suggests that the percentage of mJy and
sub-mJy radio sources that are powered by black-hole accretion
processes in this Bo\"otes field is, at least, 15$\pm$5\%.  Only 3 {\it
  sub-mJy} WSRT sources (VLBI J142934.7033, J143002.5631, and
J142835.5389) are detected out of a total of 40 potential {\it sub-mJy}
targets, implying a detection rate of 8$^{+4}_{-5}$\%.  The detection
rate is higher for {\it mJy} WSRT sources: 6 sources are detected out
of a total of 21 potential {\it mJy} targets, for a detection rate of
29$^{+11}_{-12}$\%.

The disappointing detection rate in the sub-mJy radio source population
serves to reinforce the rapid evolution that is believed to be taking
place in the radio source population at these faint mJy and sub-mJy
flux density levels \citep{mux04}.  However, the detection rate
reported here is probably further affected by resolution effects.
Several of the sub-mJy sources in our sample are likely to be
partially, rather than fully resolved.  In these cases, compact
structure associated with an active nucleus may be missed by our survey
if it falls below our 6 $\sigma$ detection threshold.  Our detection
rate of about 8\% for sub-mJy sources is, therefore, very much a
lower-limit on the fraction of sub-mJy sources with active nuclei.

The morphology of the sources detected also confirms previously
observed trends for brighter sources to exhibit radio structure on
much larger scales than fainter sources.  This can be understood in
terms of synchrotron-self-absorption theory, wherein the overall radio
source size is proportional to the square root of the peak flux
density \citep[e.g.,][]{sne00}.

\section{Future Prospects}
\label{fut}

EVN observations of the HDF-N \citep{gar01}, plus the new VLBA+GBT VLBI
observations presented here, demonstrate that deep, wide-field VLBI
surveys can now be conducted on largely random areas of sky.  In
particular, at the single-figure r.m.s. noise levels demonstrated here,
many sources can be detected simultaneously within the primary beam of
individual VLBI antennas.  This opens up the possibility of embarking
on large-area, unbiased, high-resolution surveys of the sub-mJy and mJy
radio source population.  In addition, at frequencies of a few GHz, the
combined response of the sources detected in such wide-field analyses
should always be sufficient to permit self-calibration techniques to be
utilized, irrespective of where the VLBI antennas are pointing.

Further important technical developments in this field are imminent.
Fast data output rates are being implemented in the EVN correlator at
JIVE and will permit wide-field, global VLBI studies to be conducted
with reduced losses associated with time and bandwidth smearing. The
resulting data sets will be extrememly large, typically
$\sim$Terrabyte. Fortunately the basic calibration data (e.g. phase
corrections via external calibrators) will be possible via averaged
(smaller) data sets. The full-beam calibration technique however,
requires processing of the full (un-averaged) data set and would
currently require the data to be split-up into smaller chunks
(time-ranges) and the self-calibration process spread over multiple
computer processors. A similar requirement is necessary for target
imaging, though in this case the data can be split-up both in the time
\& frequency domain. Such parallel data processing is well-suited to
emerging computer hardware, such as Linux-based PC clusters. 

These developments, together with the wide-spread adoption of the
Mark~5 recording system by the major VLBI networks \citep{whi03},
should permit global, wide-field surveys to reach 1 $\sigma$ rms noise
levels of $\sim 1~\mu$Jy~beam$^{-1}$.  At these levels of sensitivity,
a much more comprehensive census of active galaxies associated with
sub-mJy radio sources will be possible, including studies of the
optically faint microJy radio source population.  At microJy noise
levels, radio-loud active galaxies are detectable at the very earliest
cosmic epochs, when the first active galaxies and their energising
massive black holes began to form.  Certainly with these new
capabilities in place, it will be very interesting to see if the
apparent low detection rate of sub-mJy radio sources continues to be
observed at $\sim 1~\mu$Jy~beam$^{-1}$ noise levels.  Such an effect
might be expected if black holes less massive than $10^{7}$M$_{\odot}$
are rare and/or produce less radio emission than more massive systems
\citep{hai04}.

\acknowledgments This work made use of images provided by the NDWFS,
which is supported by NOAO.  NOAO is operated by the Association of
Universities for Research in Astronomy, Inc., under a cooperative
agreement with the National Science Foundation (NSF).  NRAO is a
facility of the NSF operated under cooperative agreement by Associated
Universities, Inc. The WSRT is operated by the ASTRON (Netherlands
Foundation for Research in Astronomy) with support from the
Netherlands Foundation for Scientific Research (NWO).

\clearpage

\begin{figure}
\includegraphics[angle=0,width=14cm]{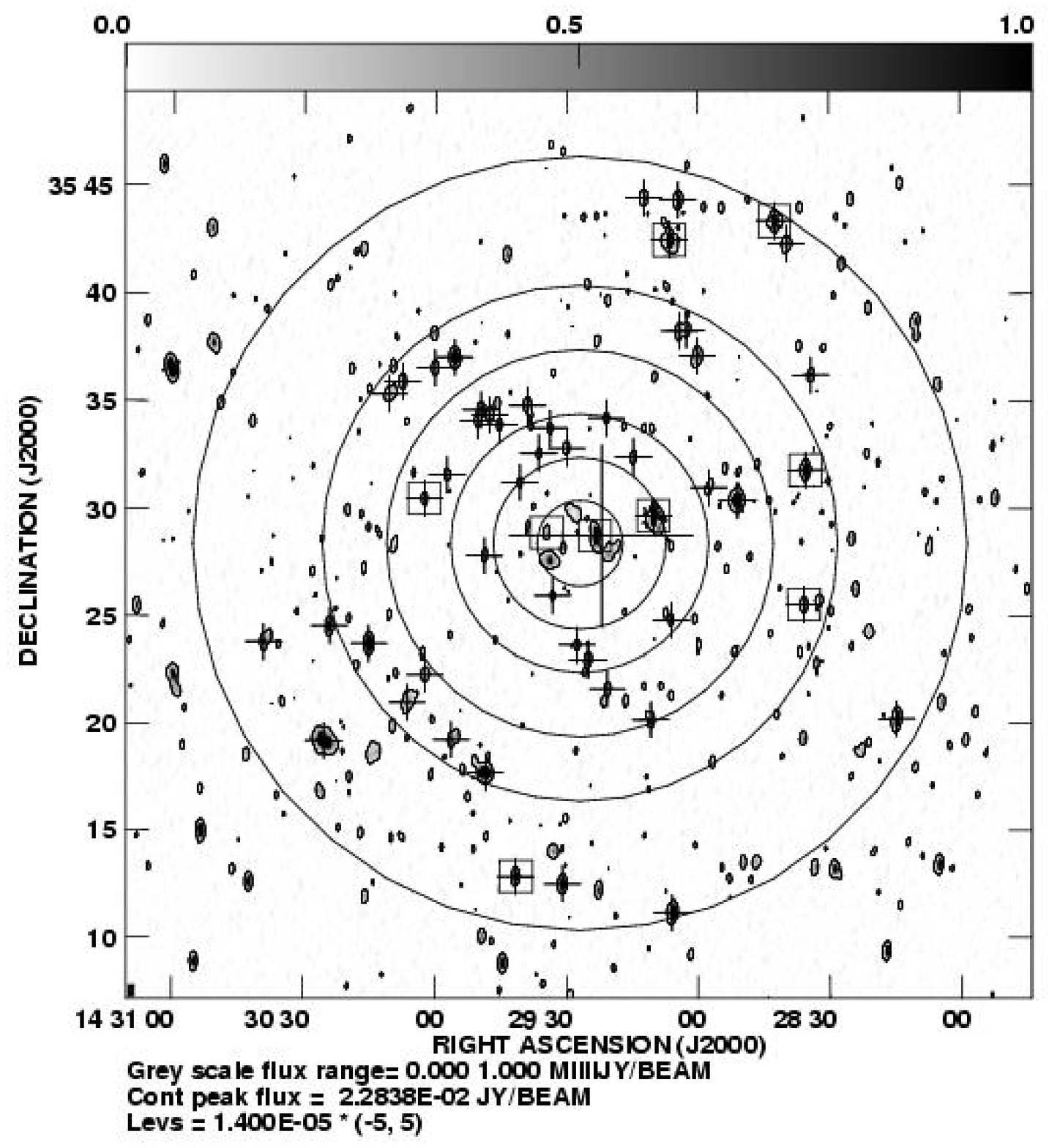}
\caption{Grey-scale WSRT image within the Bo\"otes field surveyed by
  the VLBA and GBT VLBI observations. The circles separate the various
  fields surveyed (Table~1) and are co-located on the phase-centre of
  the VLBI observations. The large cross shows the pointing position
  of the GBT (and VLBA). The extent of this cross corresponds to the
  FWHM of the GBT primary beam response. The smaller crosses located
  across the image denote regions of WSRT emission that were targeted
  by the VLBI observations. For clarity, crosses are not used to
  identify target sources within the deep and crowded 0-2\arcmin\,
  survey field. In this field all sources visible in this image were
  selected as target sources.  Targets that are also boxed identify
  those sources detected by VLBI.}
\end{figure}
\clearpage

\begin{figure}
\includegraphics[angle=-90,width=14cm]{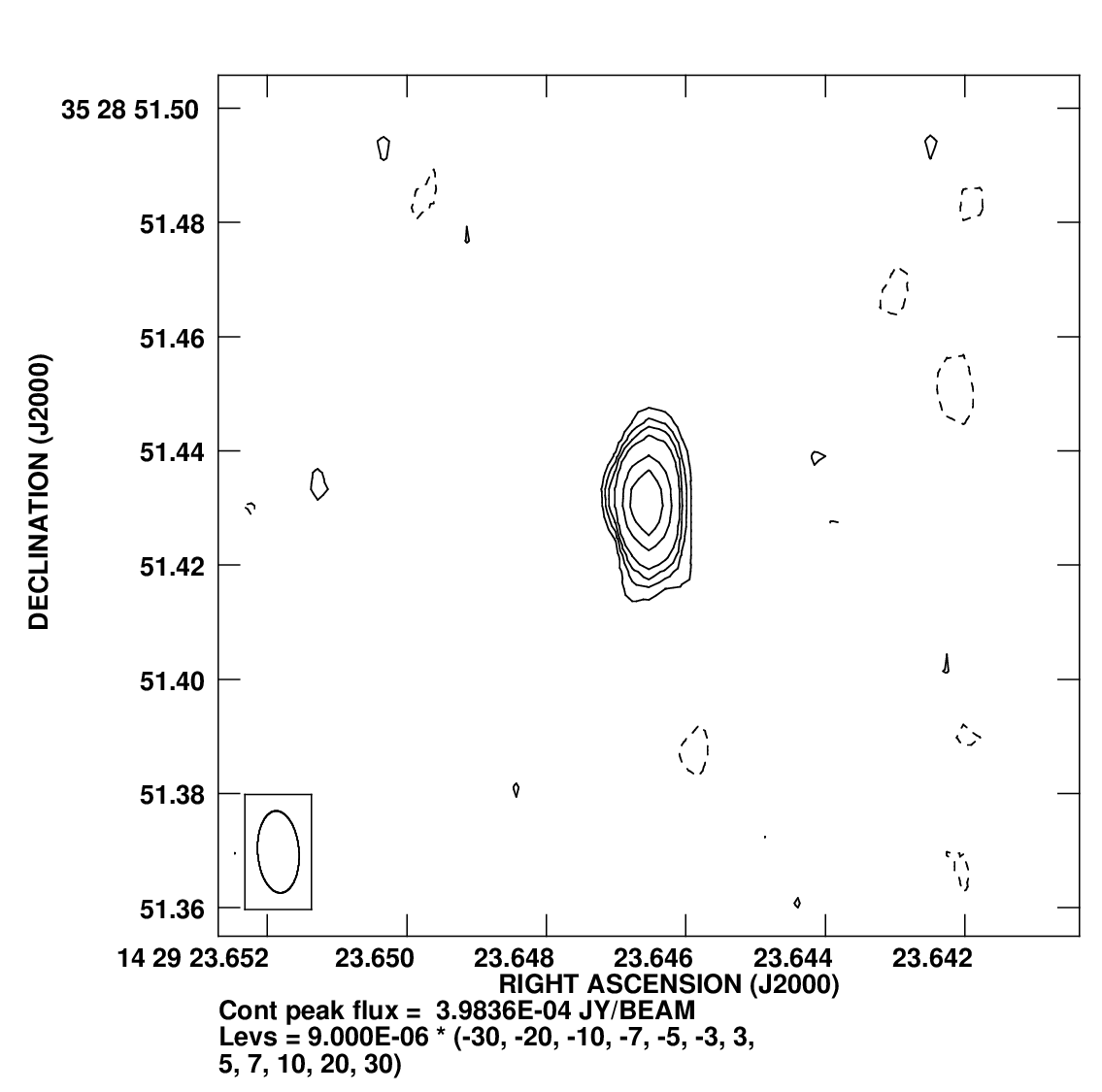}
\caption{Naturally weighted VLBI images of VLBI J142923.6466 (Target 3)
  detected in the 0-2\arcmin\, survey field. The Gaussian restoring
  beam at FWHM is $14\times7$ mas in PA=$4^{\circ}$, and is shown in the
  bottom left-hand corner of the image.  Contours are drawn at $\pm$3,
  $\pm$5, $\pm$7, $\pm$10, $\pm$20, and $\pm$30 times the 1 $\sigma$
  rms noise level of $9~\mu$Jy~beam$^{-1}$.}
\end{figure}
\clearpage

\begin{figure}
\includegraphics[angle=-90,width=14cm]{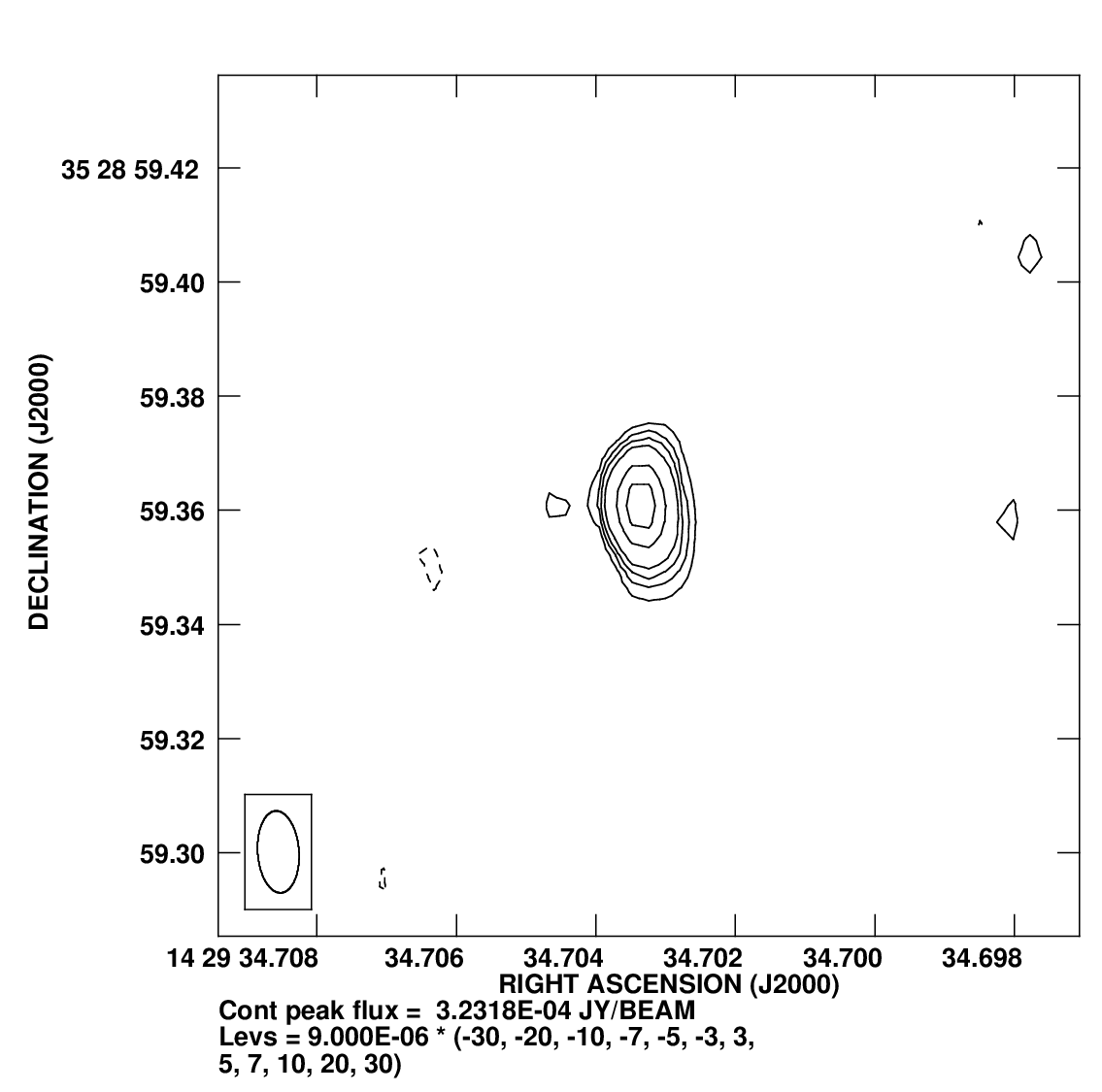}
\caption{Naturally weighted VLBI images of VLBI J142934.7033 (Target
  10) detected in the 0-2\arcmin\, survey field. The Gaussian restoring
  beam at FWHM is $14\times7$ mas in PA=$4^{\circ}$, and is shown in the
  bottom left-hand corner of the image.  Contours are drawn at $\pm$3,
  $\pm$5, $\pm$7, $\pm$10, $\pm$20, and $\pm$30 times the 1 $\sigma$
  rms noise level of $9~\mu$Jy~beam$^{-1}$.}
\end{figure}
\clearpage

\begin{figure}
\includegraphics[angle=-90,width=14cm]{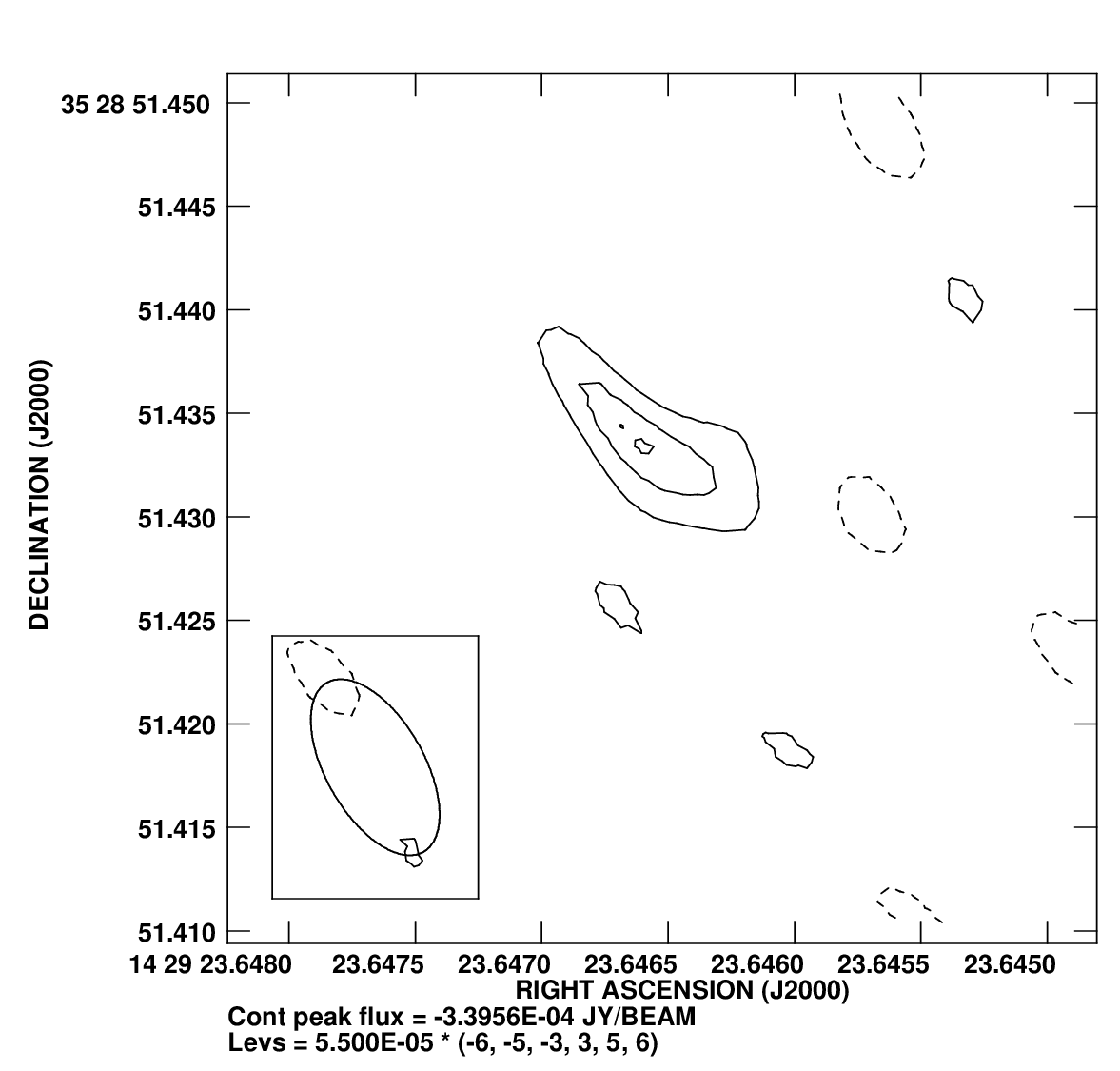}
\caption{VLBI images with uniform weighting with a robustness of -4 of
  VLBI J142923.6466 (Target 3) detected in the 0-2\arcmin\, survey
  field. The Gaussian restoring beam at FWHM is $9\times5$ mas in 
  PA=$30^{\circ}$, and is shown in the bottom left-hand corner of the
  image.  Contours are drawn at $\pm$3, $\pm$4, and $\pm$5 times the 1
  $\sigma$ rms noise level of $\sim 55~\mu$Jy~beam$^{-1}$.}
\end{figure}
\clearpage

\begin{figure}
\includegraphics[angle=-90,width=14cm]{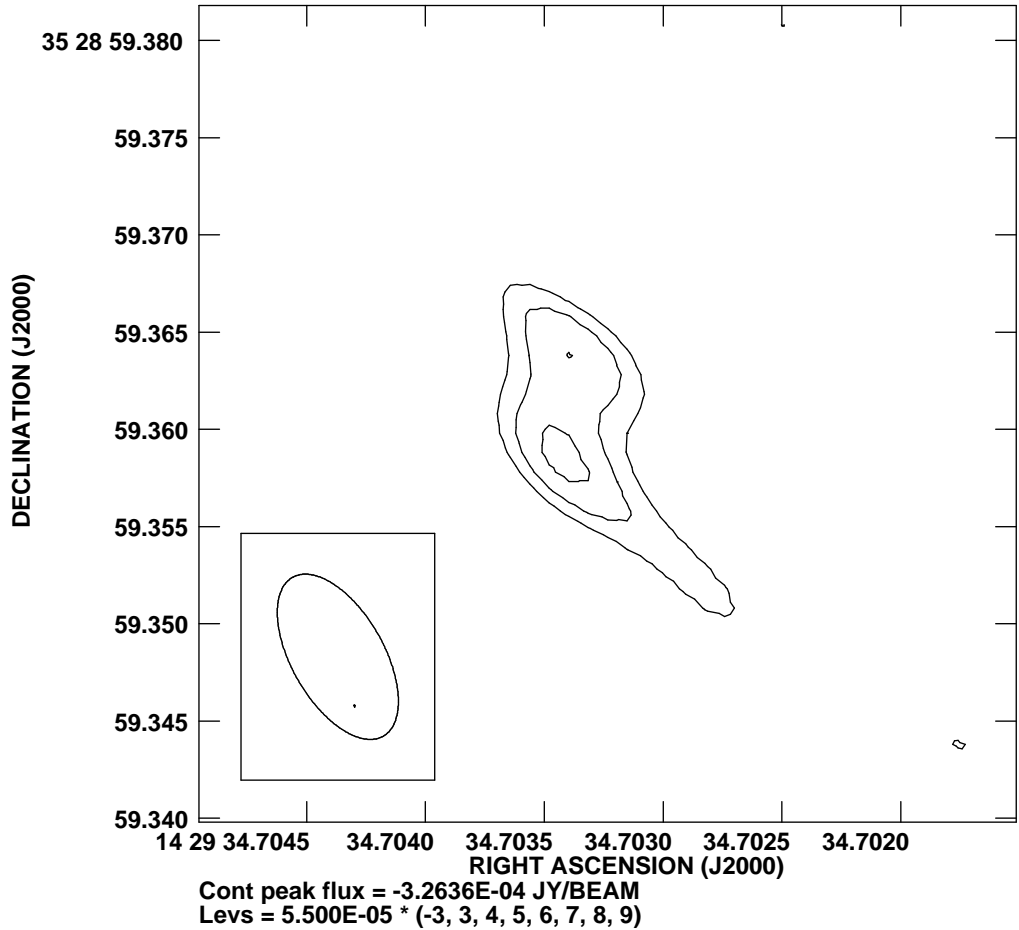}
\caption{VLBI images with uniform weighting with a robustness of -4 of
  VLBI J142934.7033 (Target 10) detected in the 0-2\arcmin\, survey
  field.  The Gaussian restoring beam at FWHM is $9\times5$ mas in 
  PA=$30^{\circ}$, and is shown in the bottom left-hand corner of the 
  image. Contours are drawn at $\pm$3, $\pm$4, and $\pm$5 times the 1
  $\sigma$ r.m.s. noise level of $\sim 55~\mu$Jy~beam$^{-1}$.}
\end{figure}
\clearpage

\begin{figure}
\includegraphics[angle=0,width=14cm]{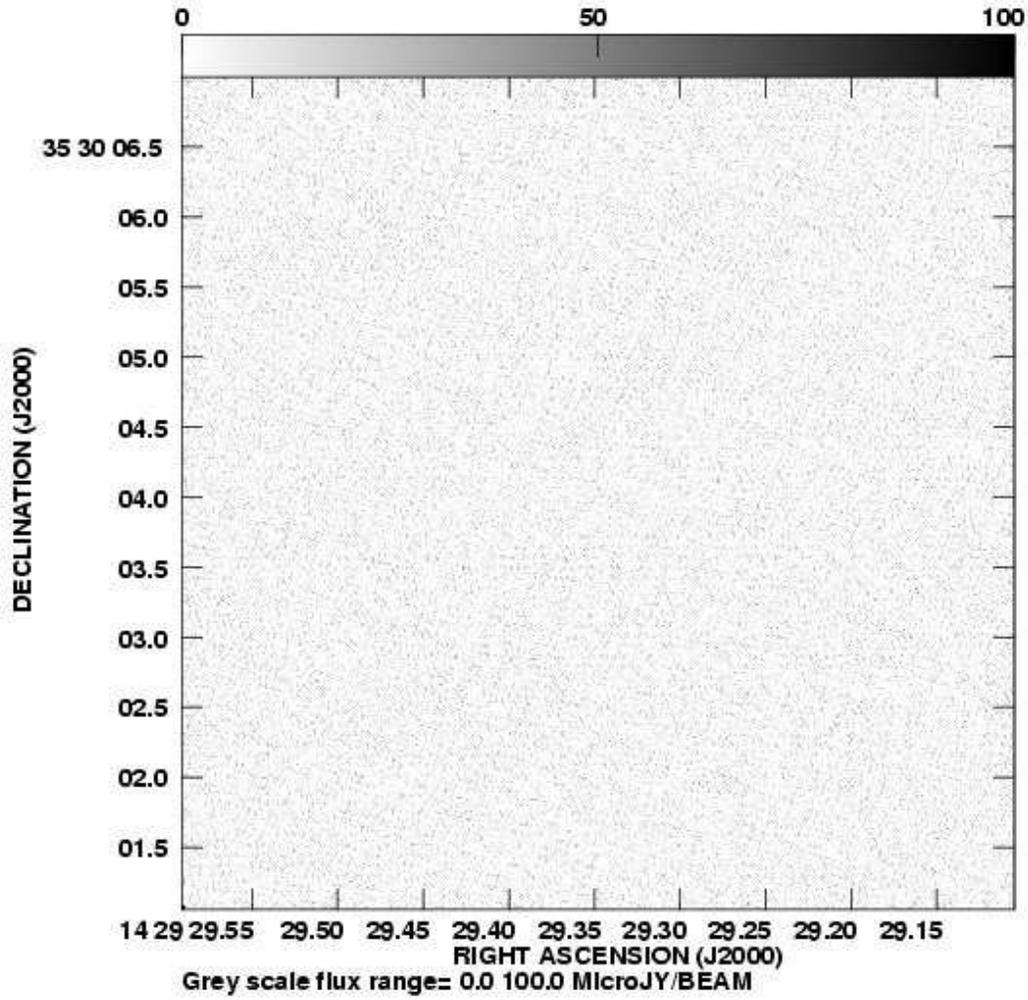}
\caption{Naturally weighted image of the (empty) region
  centered on the star forming galaxy NGC\,5646 (Target 9).  No sources
  are detected above a flux density limit of 54~$\mu$Jy~beam$^{-1}$ (the 6
  $\sigma$ r.m.s. noise level).}
\end{figure}
\clearpage

\begin{figure}
\includegraphics[angle=0,width=14cm]{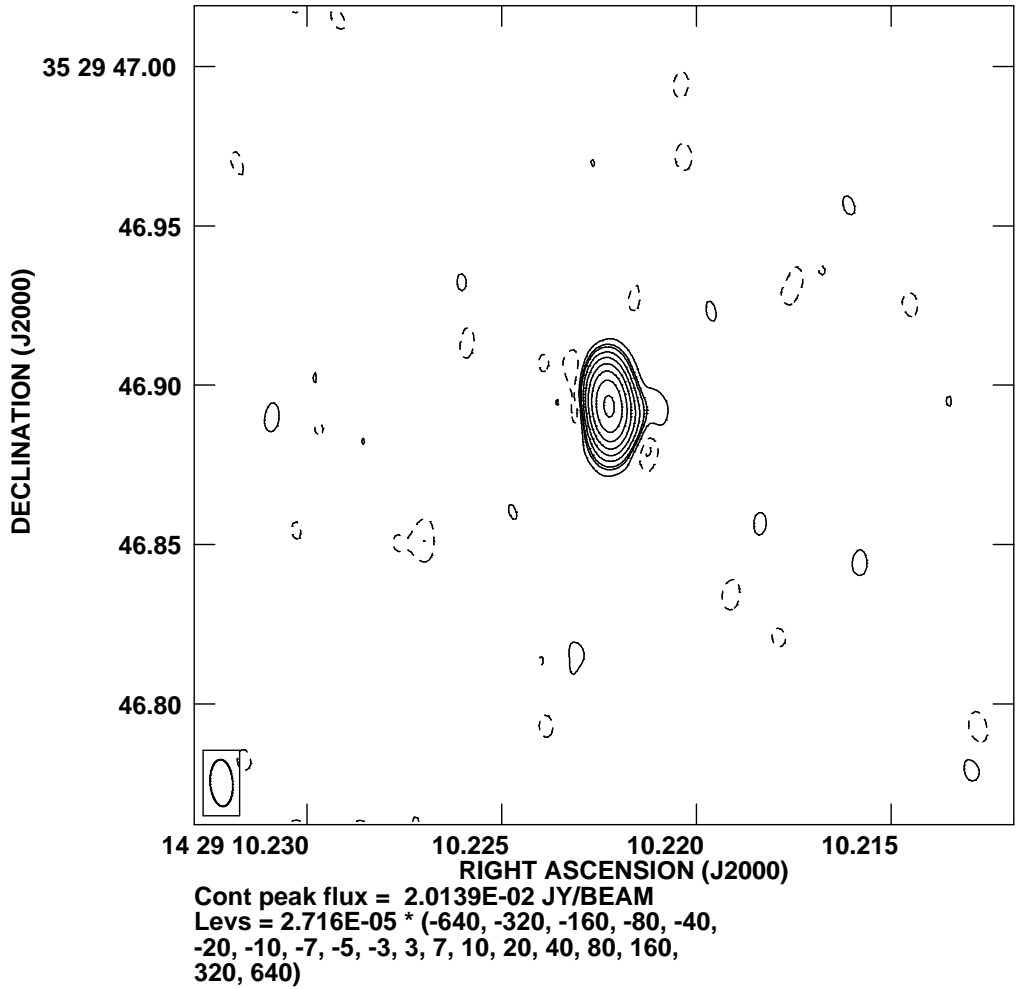}
\caption{Naturally weighted VLBI image
  of the in-beam calibrator VLBI J142910.2224 (Target 11).  Contours
  are drawn at $\pm$3, $\pm$5, $\pm$7,$\pm$10, $\pm$20....  $\pm$320
  times the 1$\sigma$ rms noise level of $\sim 27~\mu$Jy~beam$^{-1}$.
  The Gaussian restoring beam at FWHM is $14\times7$ mas in
  PA=$4^{\circ}$ shown in the bottom left-hand corner of the image.}
\end{figure}
\clearpage

\begin{figure}
\includegraphics[angle=0,width=14cm]{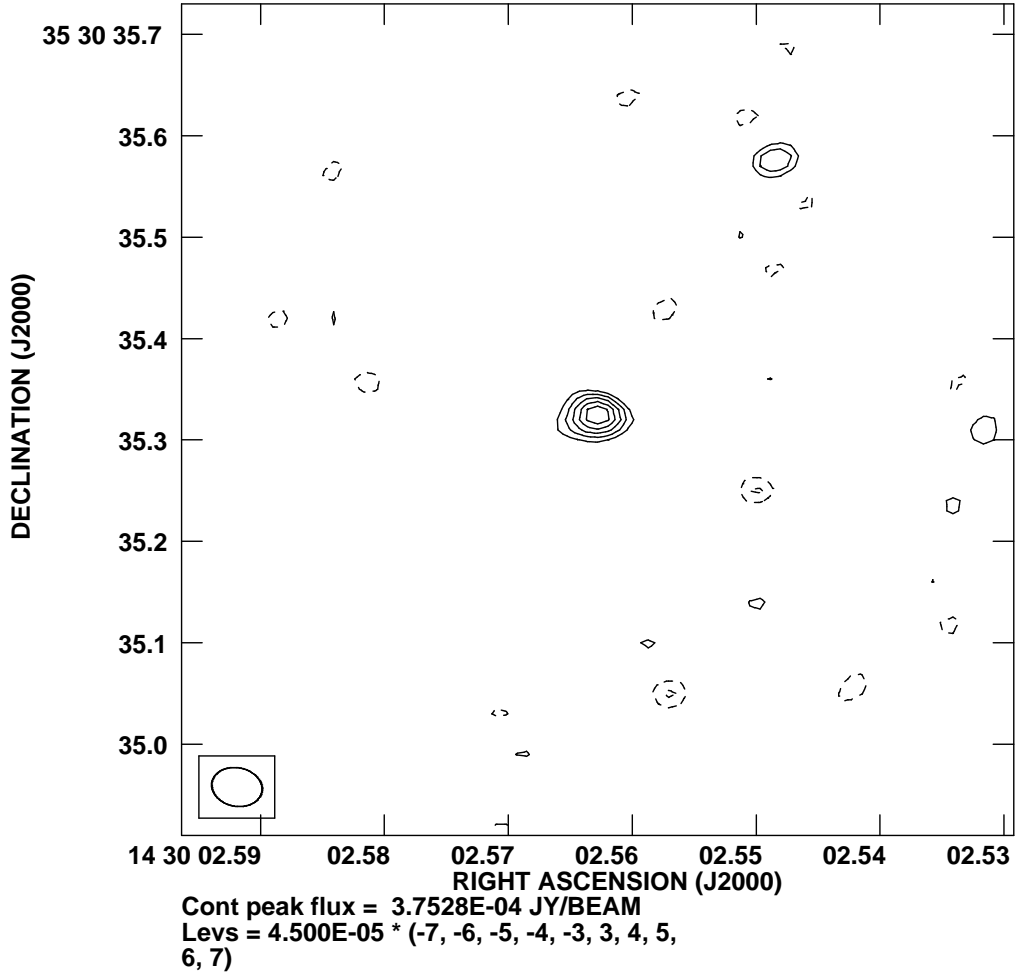}
\caption{Naturally weighted VLBI image of the only source detected
  within the 6-9\arcmin\, field, VLBI J143002.5631 (Target 33). The
  Gaussian restoring beam at FWHM is $50\times38$ mas in
  PA=$79^{\circ}$, and is shown in the bottom left-hand corner of the
  image.  Contours are drawn at $\pm$3, $\pm$4, $\pm$5, $\pm$6 and
  $\pm$7 times the 1 $\sigma$ rms noise level of $\sim
  45~\mu$Jy~beam$^{-1}$.}
\end{figure}
\clearpage

\begin{figure}
\includegraphics[angle=0,width=14cm]{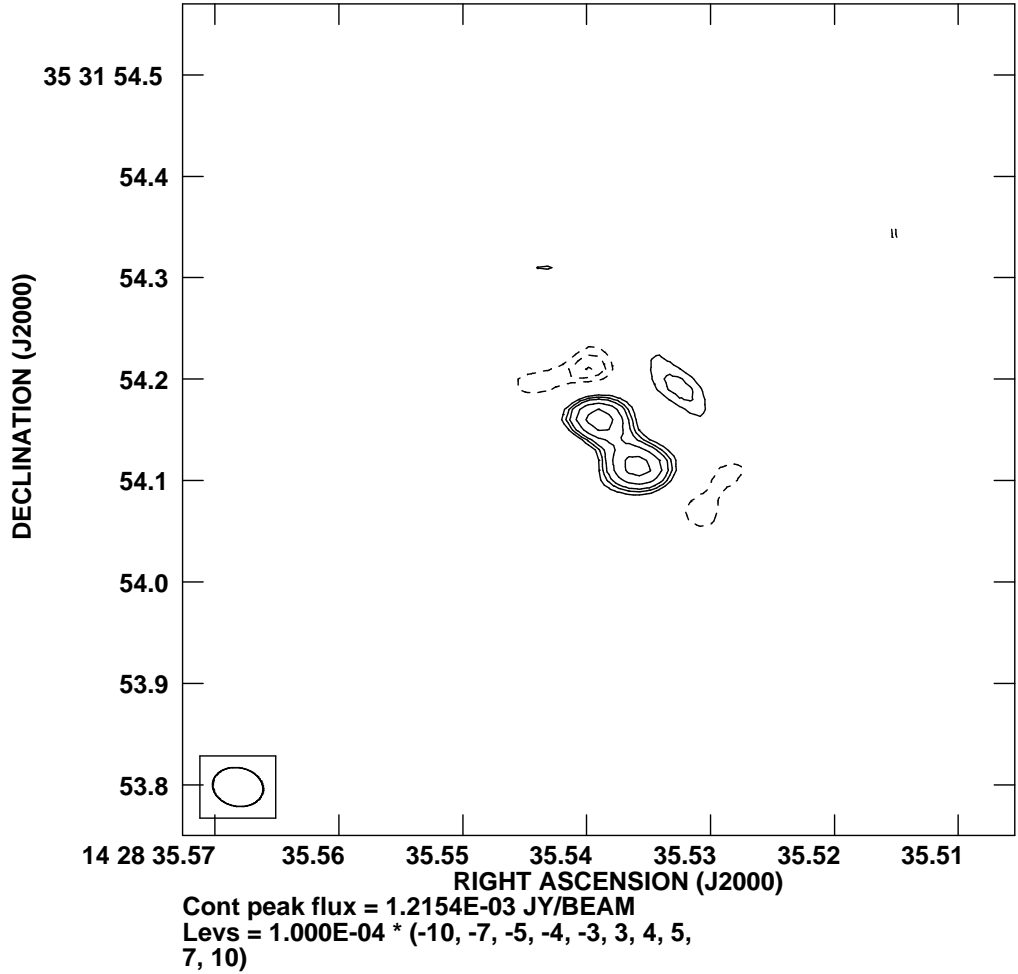}
\caption{Naturally weighted VLBI image of one of the sources detected
  within the 9-12\arcmin\, field, VLBI J142835.5389 (Target 34). The
  Gaussian restoring beam at FWHM is $50\times38$ mas in
  PA=$78^{\circ}$, and is shown in the bottom left-hand corner of the
  image.  Contours are drawn at $\pm$3, $\pm$4, $\pm$5, $\pm$7, and
  $\pm$10 times the 1 $\sigma$ (dynamic range limited) rms noise level
  of $\sim 100~\mu$Jy~beam$^{-1}$.  }
\end{figure}
\clearpage

\begin{figure}
\includegraphics[angle=0,width=14cm]{f10.ps}
\caption{Naturally weighted VLBI image of one of the sources detected
  within the 9-12\arcmin\, field, VLBI J142835.9570 (Target 35).  The
  circular FWHM Gaussian restoring beam is $50\times38$ mas in
  PA=$78^{\circ}$, and is shown in the bottom left-hand corner of the
  image.  Contours are drawn at $\pm$3, $\pm$5, and $\pm$6 times the 1
  $\sigma$ rms noise level of $\sim 55~\mu$Jy~beam$^{-1}$.  }
\end{figure}
\clearpage

\begin{figure}
\includegraphics[angle=0,width=14cm]{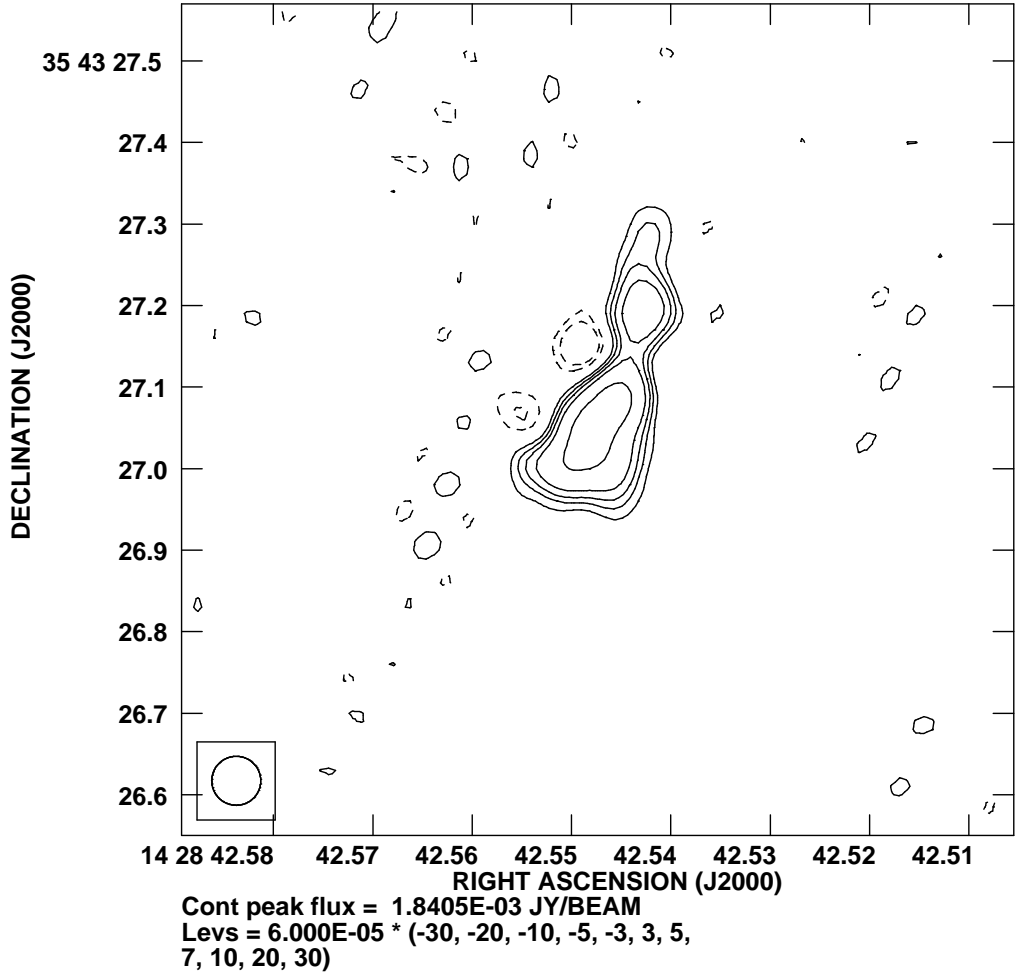}
\caption{Naturally weighted VLBI image of one of the sources detected
  within the 12-18\arcmin\, field, VLBI J142842.5476 (Target 52).  The
  circular Gaussian restoring beam at FWHM is shown in the bottom
  left-hand corner of the image.  Contours are drawn at $\pm$3,
  $\pm$5, $\pm$7, $\pm$10, $\pm$20, and $\pm$30 times the 1 $\sigma$
  rms noise level of $\sim 60~\mu$Jy~beam$^{-1}$.}
\end{figure}
\clearpage

\begin{figure}
\includegraphics[angle=0,width=14cm]{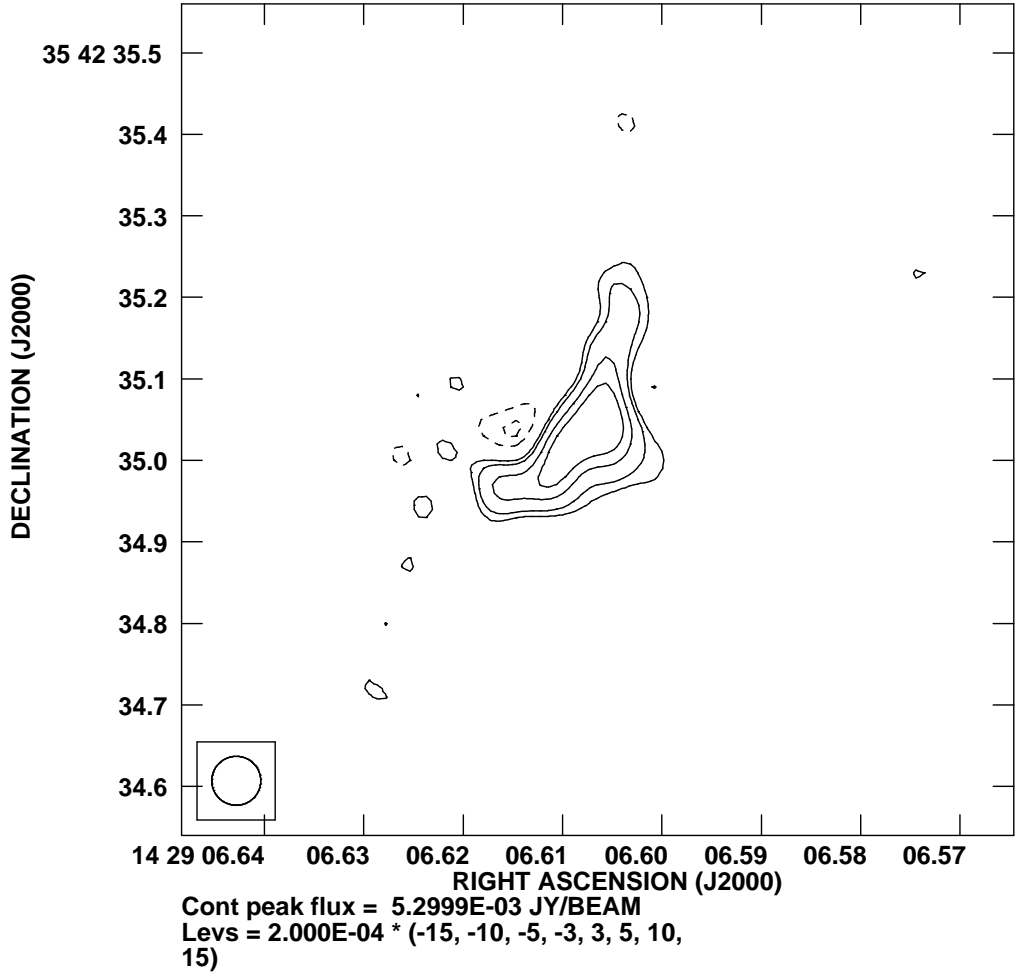}
\caption{Naturally weighted VLBI image of one of the sources detected
  within the 12-18\arcmin\, field, VLBI J142906.6095 (Target 55). The
  circular Gaussian restoring beam is 60 mas at FWHM, and is shown in
  the bottom left-hand corner of the image.  Contours are drawn at
  $\pm$3, $\pm$5, $\pm$10, and $\pm$15 times the 1 $\sigma$ rms noise
  level $\sim 200~\mu$Jy~beam$^{-1}$.}
\end{figure}
\clearpage

\begin{figure}
\includegraphics[angle=-90,width=14cm]{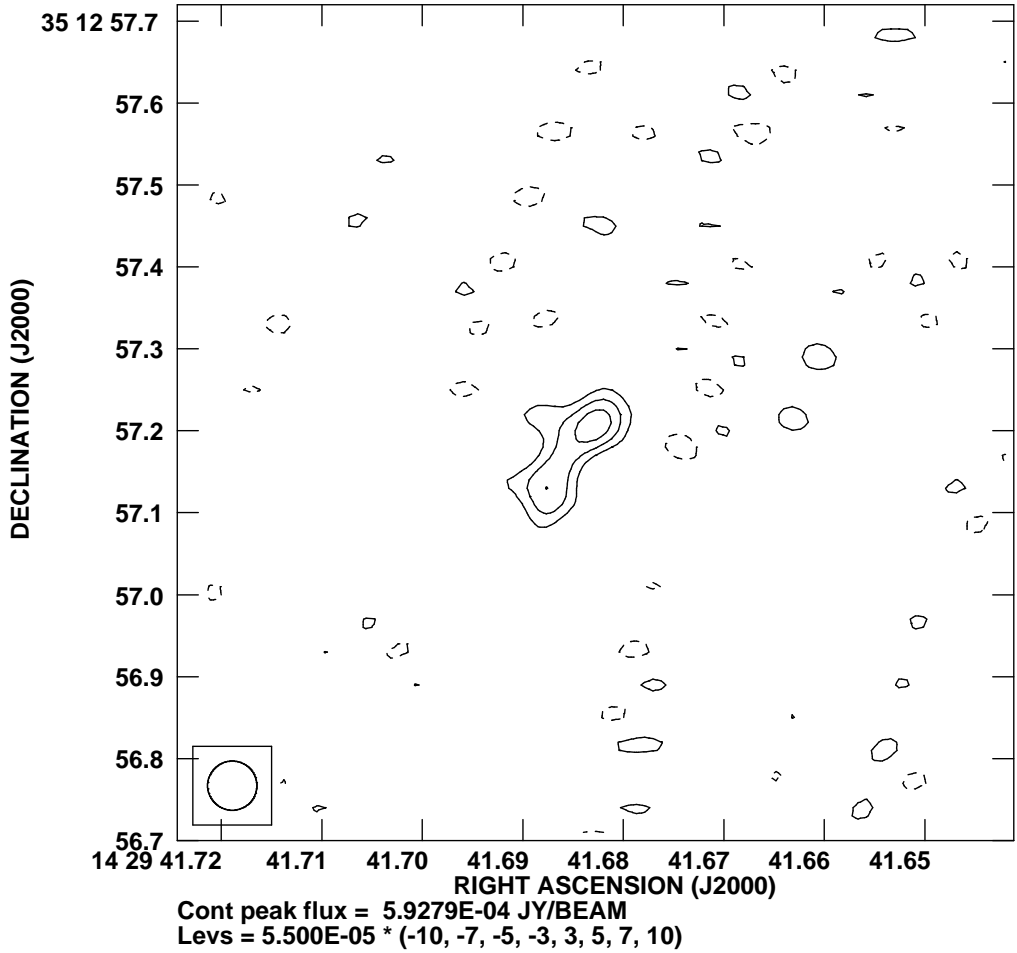}
\caption{Naturally weighted VLBI image of one of the sources detected
  within the 12-18\arcmin\, field, VLBI J142941.6843 (Target 58). The
  circular Gaussian restoring beam at FWHM is 60 mas and is shown in the bottom
  left-hand corner of the image. Contours are drawn at $\pm$3, $\pm$5,
  $\pm$7, and $\pm$10 times the 1 $\sigma$ rms noise level of $\sim
  55~\mu$Jy~beam$^{-1}$.}
\end{figure}
\clearpage

\begin{deluxetable}{ccccccc}
\tablecolumns{7}
\tabletypesize{\scriptsize}
\tablecaption{Survey Fields, Survey Depths, and Source Selection 
at 1.4~GHz}
\label{tab1}
\tablenum{1}
\tablewidth{0pc}
\tablehead{
\colhead{ }                    &
\colhead{Maximum}              &
\colhead{Maximum}              &
\colhead{ }                    &
\colhead{1 $\sigma$ rms }      &
\colhead{Survey }              &
\colhead{ }                    \\
\colhead{Survey}               &
\colhead{Radial Distance}      &
\colhead{{\em (u,v)} Range}    &
\colhead{Response}             &
\colhead{Noise}                &
\colhead{Resolution}           &
\colhead{$S_{\rm P}(WSRT)$}    \\
\colhead{Field}                &
\colhead{(\arcmin)}            &
\colhead{(M$\lambda$)}         &
\colhead{$R$}                  &
\colhead{($\mu$Jy~beam$^{-1}$)}&
\colhead{(mas, mas)}           &
\colhead{(mJy~beam$^{-1}$)}    \\
\colhead{(1)}& \colhead{(2)}& \colhead{(3)}& \colhead{(4)}&
\colhead{(5)}& \colhead{(6)}& \colhead{(7)}\\
}
\startdata
  0-2\arcmin &  2 & 24 & 0.73 &  9 & $14  \times  7$ & $> 0.074$ \\
  2-4\arcmin &  4 & 12 & 0.47 & 11 & $18  \times  9$ & $> 0.14$  \\
  4-6\arcmin &  6 &  8 & 0.75 & 19 & $26  \times 20$ & $> 0.23$  \\
  6-9\arcmin &  9 &  6 & 0.62 & 37 & $33  \times 27$ & $> 0.36$  \\
 9-12\arcmin & 12 &  4 & 0.52 & 45 & $115 \times 33$ & $> 0.50$  \\
12-18\arcmin & 18 &  3 & 0.27 & 55 & $104 \times 38$ & $> 1.2$   \\
\enddata
\end{deluxetable}

\begin{deluxetable}{cllccccc}
\tablecolumns{8}
\tabletypesize{\scriptsize}
\tablecaption{Astrometry and Photometry at 1.4~GHz for the
0-2\arcmin\, Field}
\label{tab2}
\tablenum{2}
\tablewidth{0pc}
\tablehead{ \colhead{                   }&
            \colhead{ Right Ascension   }&
            \colhead{ Declination       }&
            \colhead{ $S_{\rm P}$       }&
            \colhead{ $S_{\rm I}$       }&
            \colhead{ $LAS$             }\\
            \colhead{ Target            }&
            \colhead{ (J2000)           }&
            \colhead{ (J2000)           }&
            \colhead{ (mJy~beam$^{-1}$) }&
            \colhead{ (mJy)             }&
            \colhead{ ($\arcsec$)       }\\
\colhead{(1)}& \colhead{(2)}& \colhead{(3)}& \colhead{(4)}&
\colhead{(5)}& \colhead{(6)}\\
}
\startdata
1  & 14 29 18.545  & $+$35 28 21.04   & $0.202\pm0.014$ &
     $0.202\pm0.014$ & \nodata        \\

2  & 14 29 20.493  & $+$35 28 00.79    & $0.537 \pm 0.013$ &
     $1.21 \pm 0.04$ & $< 20 $         \\

3  & 14 29 23.541  & $+$35 28 52.08   & $3.92 \pm 0.013$ &
     $6.053 \pm 0.031$ & $< 25 $      \\
   & 14 29 23.6466 & $+$35 28 51.431 & 0.398$\pm$0.009 &
     0.461$\pm$0.018 & $<$0.008      \\

4  & 14 29 25.367  & $+$35 29 38.73   & $0.110 \pm 0.014$ &
     $0.164 \pm 0.031$ & $ < 17 $     \\

5  & 14 29 27.600  & $+$35 29 50.68    & $0.219 \pm 0.013$ &
     $0.776 \pm 0.057$ & $ < 39 $      \\

6  & 14 29 27.627  & $+$35 29 03.00    & $0.105 \pm 0.014 $ &
     $0.138 \pm 0.028$ & $< 25  $      \\

7  & 14 29 29.348  & $+$35 30 04.03   & $0.225 \pm 0.014 $ &
     $0.267 \pm 0.027$ & $< 11  $     \\

8  & 14 29 30.865  & $+$35 28 16.57   & $0.237 \pm 0.014 $ &
     $0.283 \pm 0.027$ & $< 14  $     \\

9  & 14 29 33.952  & $+$35 27 42.55   & $1.128 \pm 0.014 $ &
     $1.702 \pm 0.031$ & $< 12  $     \\

10 & 14 29 34.632  & $+$35 28 59.50   & $0.571 \pm 0.014 $ &
     $0.663 \pm 0.026$ & $< 10  $     \\
   & 14 29 34.7033 & $+$35 28 59.361 & 0.344$\pm$0.009 &
     0.455$\pm$0.019 & $<$0.010      \\
     
     \enddata \tablecomments{Units of right ascension are hours,
       minutes, and seconds, and units of declination are degrees,
       arcminutes, and arcseconds. The first data entry for each target
       source refers to the WSRT image and the second entry,
       when present, refers to the VLBI image.  }

\end{deluxetable}
\clearpage

\begin{deluxetable}{cllccccc}
\tablecolumns{8}
\tabletypesize{\scriptsize}
\tablecaption{Astrometry and Photometry at 1.4~GHz for the
2-4\arcmin\, Field}
\label{tab3}
\tablenum{3}
\tablewidth{0pc}
\tablehead{ \colhead{                   }&
            \colhead{ Right Ascension   }&
            \colhead{ Declination       }&
            \colhead{ $S_{\rm P}$       }&
            \colhead{ $S_{\rm I}$       }&
            \colhead{ $LAS$             }\\
            \colhead{ Target            }&
            \colhead{ (J2000)           }&
            \colhead{ (J2000)           }&
            \colhead{ (mJy~beam$^{-1}$) }&
            \colhead{ (mJy)             }&
            \colhead{ ($\arcsec$)       }\\
\colhead{(1)}& \colhead{(2)}& \colhead{(3)}& \colhead{(4)}&
\colhead{(5)}& \colhead{(6)}\\
}
\startdata

11 & 14 29 10.130  & $+$35 29 46.13   & $23.337 \pm 0.014$ &
     $25.099 \pm 0.025$ & $< 5 $      \\
   & 14 29 10.2224 & $+$35 29 46.893  & $20.054\pm0.013$ &
     $21.078\pm$0.024  & $ < 0.003$   \\

12 & 14 29 33.273 & $+$35 26 03.20    & $ 0.184\pm0.014 $ &
     $ 0.190\pm 0.025$     & $<18  $  \\

13 & 14 29 40.683 & $+$35 31 20.15    & $0.167 \pm 0.014 $ &
     $ 0.233\pm0.030 $  & $< 14 $     \\

\enddata

\tablecomments{Units of right ascension are hours, minutes, and
  seconds, and units of declination are degrees, arcminutes, and
  arcseconds. The first data entry for each target source refers to the
  WSRT image and the second entry, when present, refers to the VLBI
  image. }

\end{deluxetable}
\clearpage

\begin{deluxetable}{cllccccc}
\tablecolumns{8}
\tabletypesize{\scriptsize}
\tablecaption{Astrometry and Photometry at 1.4~GHz for the
4-6\arcmin\, Field}
\tablewidth{0pc}
\label{tab4}
\tablenum{4}
\tablehead{ \colhead{                   }&
            \colhead{ Right Ascension   }&
            \colhead{ Declination       }&
            \colhead{ $S_{\rm P}$       }&
            \colhead{ $S_{\rm I}$       }&
            \colhead{ $LAS$             }\\
            \colhead{ Target            }&
            \colhead{ (J2000)           }&
            \colhead{ (J2000)           }&
            \colhead{ (mJy~beam$^{-1}$) }&
            \colhead{ (mJy)             }&
            \colhead{ ($\arcsec$)       }\\
\colhead{(1)}& \colhead{(2)}& \colhead{(3)}& \colhead{(4)}&
\colhead{(5)}& \colhead{(6)}\\}
\startdata

14 & 14 29 06.132 & $+$35 24 55.09 & $0.290\pm0.014 $ &
     $0.290\pm0.014$  & \nodata    \\

15 & 14 29 14.758 & $+$35 32 31.50 & $ 0.364\pm0.014 $ &
     $ 0.419\pm0.026 $   & $<13  $ \\

16 & 14 29 20.941 & $+$35 34 18.13 & $ 0.314\pm0.014 $ &
     $0.314\pm0.014 $ & \nodata    \\

17 & 14 29 25.171 & $+$35 23 04.67 & $ 0.893\pm0.026 $ &
     $0.801 \pm 0.014$   & $< 10 $ \\

18 & 14 29 25.178 & $+$35 23 04.91 & $0.426 \pm 0.014$ &
     $0.426 \pm 0.023$   & $< 7 $  \\

19 & 14 29 27.698 & $+$35 23 45.31 & $ 0.237\pm0.014 $ &
     $0.237\pm0.014 $   & \nodata  \\

20 & 14 29 30.045 & $+$35 32 55.74 & $ 0.432\pm0.014 $ &
     $0.432\pm0.014  $   & \nodata \\

21 & 14 29 33.869 & $+$35 33 50.16 & $ 0.267\pm0.023 $ &
     $0.290\pm0.014 $   & $<9  $   \\

22 & 14 29 36.400 & $+$35 32 41.50 & $0.243 \pm0.014 $ &
     $0.243 \pm0.014 $  & \nodata  \\

23 & 14 29 45.319 & $+$35 33 59.07 & $0.401 \pm 0.015$ &
     $0.401\pm0.025$ & $ < 7 $     \\

24 & 14 29 48.777 & $+$35 27 55.01 & $ 0.428\pm0.025 $ &
     $0.415 \pm0.014 $   & $<11  $ \\

\enddata

\tablecomments{Units of right ascension are hours, minutes, and
  seconds, and units of declination are degrees, arcminutes, and
  arcseconds. The first data entry for each target source refers to the
  WSRT image and the second entry, when present, refers to the VLBI
  image. }

\end{deluxetable}
\clearpage

\begin{deluxetable}{cllccccc}
\tablecolumns{8}
\tabletypesize{\scriptsize}
\tablecaption{Astrometry and Photometry at 1.4~GHz for the
6-9\arcmin\, Field}
\tablewidth{0pc}
\label{tab5}
\tablenum{5}
\tablehead{ \colhead{                   }&
            \colhead{ Right Ascension   }&
            \colhead{ Declination       }&
            \colhead{ $S_{\rm P}$       }&
            \colhead{ $S_{\rm I}$       }&
            \colhead{ $LAS$             }\\
            \colhead{ Target            }&
            \colhead{ (J2000)           }&
            \colhead{ (J2000)           }&
            \colhead{ (mJy~beam$^{-1}$) }&
            \colhead{ (mJy)             }&
            \colhead{ ($\arcsec$)       }\\
\colhead{(1)}& \colhead{(2)}& \colhead{(3)}& \colhead{(4)}&
\colhead{(5)}& \colhead{(6)}\\
}

\startdata

25 & 14 28 50.931  & $+$35 30 28.97  & $16.328 \pm 0.014$ &
     $21.333 \pm 0.028$   & $< 8 $   \\

26 & 14 29 10.769  & $+$35 20 16.94  & $0.687 \pm 0.014$ &
     $0.831 \pm 0.027$   & $< 12 $   \\

27 & 14 29 38.981  & $+$35 34 53.02  & $0.9736 \pm 0.014$ &
     $1.060 \pm 0.026$   & $< 9 $    \\

28 & 14 29 47.538  & $+$35 34 27.31  & $0.672 \pm 0.014$ &
     $0.764 \pm 0.026$   & $< 9 $    \\

29 & 14 29 49.542  & $+$35 34 41.43  & $1.121 \pm 0.014$ &
     $1.217 \pm 0.025$   & $< 7 $    \\

30 & 14 29 50.247  & $+$35 34 11.09  & $0.459 \pm 0.014$ &
     $0.459 \pm 0.014$ & \nodata     \\

31 & 14 29 57.320  & $+$35 31 41.34  & $0.377 \pm 0.014$ &
     $0.410\pm0.026$ & $ < 9 $       \\

32 & 14 29 57.428  & $+$35 31 04.90  & $0.427 \pm 0.014$ &
     $0.571\pm0.030$ & $ < 18 $      \\

33 & 14 30 02.494  & $+$35 30 34.66  & $0.420 \pm 0.014$ &
     $0.439\pm0.025$ & $ < 9 $       \\
   & 14 30 02.5631 & $+$35 30 35.323 & $0.345 \pm 0.052$ &
     $0.508 \pm 0.119$ & $<0.063 $   \\

\enddata

\tablecomments{Units of right ascension are hours, minutes, and
  seconds, and units of declination are degrees, arcminutes, and
  arcseconds. The first data entry for each target source refers to the
  WSRT image and the second entry, when present, refers to the VLBI
  image. }

\end{deluxetable}
\clearpage

\begin{deluxetable}{cllccccc}
\tablecolumns{8}
\tabletypesize{\scriptsize}
\tablecaption{Astrometry and Photometry at 1.4~GHz for the
9-12\arcmin\, Field}
\tablewidth{0pc}
\label{tab6}
\tablenum{6}
\tablehead{ \colhead{                   }&
            \colhead{ Right Ascension   }&
            \colhead{ Declination       }&
            \colhead{ $S_{\rm P}$       }&
            \colhead{ $S_{\rm I}$       }&
            \colhead{ $LAS$             }\\
            \colhead{ Target            }&
            \colhead{ (J2000)           }&
            \colhead{ (J2000)           }&
            \colhead{ (mJy~beam$^{-1}$) }&
            \colhead{ (mJy)             }&
            \colhead{ ($\arcsec$)       }\\
\colhead{(1)}& \colhead{(2)}& \colhead{(3)}& \colhead{(4)}&
\colhead{(5)}& \colhead{(6)}\\
}

\startdata

34 & 14 28 35.444  & $+$35 31 53.34   & $6.922 \pm 0.014$  &
     $7.504 \pm 0.028$   & $< 5 $     \\
   & 14 28 35.5359 & $+$35 31 54.114 & $1.208\pm0.060$  &
     $1.327\pm0.111$    & $< 0.029$  \\
   & 14 28 35.5389 & $+$35 31 54.160 & $1.250\pm0.060$  &
     $1.250\pm0.060$    & \nodata    \\

35 & 14 28 35.858  & $+$35 25 36.64   & $0.896 \pm 0.014$  &
     $0.997 \pm 0.026$   & $< 8 $     \\
   & 14 28 35.9570 & $+$35 25 37.685  & $0.430\pm0.050$  &
     $0.455\pm0.093$    & $ < 0.053$  \\

36 & 14 29 00.076  & $+$35 37 11.30   & $0.8524 \pm 0.013$  &
     $1.802 \pm 0.039$   & $< 24 $    \\

37 & 14 29 02.588  & $+$35 38 24.88   & $0.7806 \pm 0.014$  &
     $0.912 \pm 0.027$   & $< 8 $     \\

38 & 14 29 04.128  & $+$35 38 21.36   & $0.975 \pm 0.014$  &
     $1.089 \pm 0.026$   & $< 9 $     \\

39 & 14 29 20.670  & $+$35 21 43.80   & $0.451 \pm 0.014$  &
     $0.538 \pm 0.028$   & $< 11 $    \\

40 & 14 29 48.525  & $+$35 17 47.08   & $3.843 \pm 0.014$  &
     $11.571 \pm 0.052$   & $< 28 $   \\

41 & 14 29 55.571  & $+$35 37 06.92   & $9.075 \pm 0.014$  &
     $10.443 \pm 0.026$   & $< 6 $    \\

42 & 14 29 56.252  & $+$35 19 20.17   & $0.683 \pm 0.014$  &
     $0.887 \pm 0.028$   & $< 11 $    \\

43 & 14 30 00.129  & $+$35 36 38.07   & $0.6535 \pm 0.014$  &
     $0.7112 \pm 0.026$   & $< 8 $    \\

44 & 14 30 02.362  & $+$35 22 22.85   & $0.515 \pm 0.014$  &
     $6.690 \pm 0.028$   & $< 10 $    \\

45 & 14 30 06.389  & $+$35 21 04.45   & $1.019 \pm 0.014$  &
     $1.019 \pm 0.014$   & \nodata    \\

46 & 14 30 07.560  & $+$35 35 59.62   & $0.873 \pm 0.014$  &
     $0.989 \pm 0.026$   & $< 9 $     \\

47 & 14 30 10.649  & $+$35 35 25.21   & $0.697 \pm 0.014$  &
     $0.697 \pm 0.014$   & \nodata    \\

48 & 14 30 15.192  & $+$35 23 47.45   & $8.295 \pm 0.014$  &
     $8.926 \pm 0.025$   & $< 6 $     \\

\enddata

\tablecomments{Units of right ascension are hours, minutes, and
  seconds, and units of declination are degrees, arcminutes, and
  arcseconds. The first data entry for each target source refers to the
  WSRT image and the second entry, when present, refers to the VLBI
  image. }

\end{deluxetable}
\clearpage

\begin{deluxetable}{cllccccc}
\tablecolumns{8}
\tabletypesize{\scriptsize}
\tablecaption{Astrometry and Photometry at 1.4~GHz for the
12-18\arcmin\, Field}
\tablewidth{0pc}
\label{tab7}
\tablenum{7}
\tablehead{ \colhead{                   }&
            \colhead{ Right Ascension   }&
            \colhead{ Declination       }&
            \colhead{ $S_{\rm P}$       }&
            \colhead{ $S_{\rm I}$       }&
            \colhead{ $LAS$             }\\
            \colhead{ Target            }&
            \colhead{ (J2000)           }&
            \colhead{ (J2000)           }&
            \colhead{ (mJy~beam$^{-1}$) }&
            \colhead{ (mJy)             }&
            \colhead{ ($\arcsec$)       }\\
\colhead{(1)}& \colhead{(2)}& \colhead{(3)}& \colhead{(4)}&
\colhead{(5)}& \colhead{(6)}\\
}

\startdata

49 & 14 28 14.636  & $+$35 20 15.98   & $3.855 \pm 0.013$  &
     $6.040 \pm 0.032$   & $< 24 $    \\

50 & 14 28 34.192  & $+$35 36 16.86   & $0.625 \pm 0.014$  &
     $0.668 \pm 0.025$   & $< 7 $     \\

51 & 14 28 39.791  & $+$35 42 23.14   & $1.270 \pm 0.014$  &
     $1.528 \pm 0.027$   & $< 8 $     \\

52 & 14 28 42.457  & $+$35 43 26.15   & $9.379 \pm 0.014$  &
     $10.430 \pm 0.026$   & $< 5 $    \\
   & 14 28 42.5476 & $+$35 43 27.053  & $1.909\pm0.054$    &
     $5.472\pm0.203$  & $<0.181$      \\

53 & 14 29 04.519  & $+$35 44 24.60   & $2.705 \pm 0.014$  &
     $3.026 \pm 0.026$   & $< 5 $     \\

54 & 14 29 05.907  & $+$35 11 14.99   & $6.944 \pm 0.014$  &
     $8.266 \pm 0.028$   & $< 8 $     \\

55 & 14 29 06.517  & $+$35 42 34.16   & $18.504 \pm 0.014$  &
     $20.042 \pm 0.026$   & $< 5 $    \\
   & 14 29 06.6095 & $+$35 42 35.012 & $7.442\pm0.070$ &
     $13.216\pm0.182$   & $<0.109$     \\

56 & 14 29 12.314  & $+$35 44 29.88   & $1.208 \pm 0.014$  &
     $1.352 \pm 0.026$   & $< 6 $     \\

57 & 14 29 30.993  & $+$35 12 36.75   & $3.124 \pm 0.014$  &
     $3.499 \pm 0.026$   & $< 6 $     \\

58 & 14 29 41.602  & $+$35 12 56.73   & $1.852 \pm 0.014$  &
     $2.012 \pm 0.025$   & $< 7 $     \\
   & 14 29 41.6843 & $+$35 12 57.189 & $0.569\pm0.065$
     & $0.8652\pm0.151$ & $ < 0.129 $ \\

59 & 14 30 24.204  & $+$35 24 37.12   & $2.381 \pm 0.014$  &
     $2.542 \pm 0.025$   & $< 8 $     \\

60 & 14 30 25.477  & $+$35 19 15.21   & $2.441 \pm 0.013$  &
     $1.221 \pm 0.075$   & $< 40 $    \\

61 & 14 30 39.112  & $+$35 23 50.76   & $1.213 \pm 0.014$  &
     $1.432 \pm 0.027$   & $< 10 $    \\

\enddata

\tablecomments{Units of right ascension are hours, minutes, and
  seconds, and units of declination are degrees, arcminutes, and
  arcseconds. The first data entry for each target source refers to the
  WSRT image and the second entry, when present, refers to the VLBI
  image. }

\end{deluxetable}
\clearpage

\end{document}